\documentclass[a4paper,fleqn,usenatbib]{mnras}

\usepackage{newtxtext,newtxmath}

\usepackage[T1]{fontenc}
\usepackage{ae,aecompl}

\usepackage{graphicx}	
\usepackage{amsmath}	
\usepackage{amssymb}	
\usepackage{physics}
\usepackage{enumitem}
\usepackage{color}
\newcommand{\kms}{$\rm{km \,s^{-1}}$}
\newcommand{\cmtwo}{$\rm{cm^{-2}}$}

\title[Weak Winds in AGN PSBs ]{Weak Galactic Winds in Active Galactic Nuclei Post-starburst Host Galaxies at $\MakeLowercase{z} \sim 0.1$}
\author[Yesuf et al.]{
Hassen M. Yesuf \thanks{E-mail: hyesuf@ucolick.org},
S.M. Faber,
David C. Koo,
Lin Lin Lee
\\
University of California Observatories and Department of Astronomy \& Astrophysics, University of California, Santa Cruz, CA 95064, USA \\
}

\date{Accepted XXX. Received YYY; in original form ZZZ}

\pubyear{2017}

\begin{document}
\label{firstpage}
\pagerange{\pageref{firstpage}--\pageref{lastpage}}
\maketitle

\begin{abstract}
Post-starburst (PSB) galaxies may be in rapid transition from star-forming to quiescence and are excellent candidates to constrain active galactic nuclei (AGN) feedback models. We study galactic winds in the stacked spectrum of 560 AGN PSBs and that of a control sample of star-forming galaxies in the Sloan Digital Sky Survey (SDSS). Using a two component (inter-stellar +wind) absorption-line model of the \ion{Na}{i} doublet, and after accounting for the stellar photospheric absorption, we find that the AGN PSBs have a centroid wind velocity shift of $-252^{+64}_{-57}$ km\,s$^{-1}$ and a maximum blueshift velocity of $-678^{+54}_{-53}$ km\,s$^{-1}$. In comparison, the control sample, which is matched with the AGN PSBs in redshift, stellar mass, axis-ratio, the 4000\,{\AA} break index, Balmer decrement, and WISE 12\,$\mu$m to 4.6\,$\mu$m flux ratio, has a centroid wind velocity shift of  $-119^{+33}_{-41}$ km s$^{-1}$ and  a maximum velocity of $-406^{+51}_{-61}$ km\,s$^{-1}$. The equivalent widths due to the winds in both samples are similar: $0.36^{+0.10}_{-0.07}$\,{\AA} for the AGN PSBs and $0.24^{+0.07}_{-0.06} $\,{\AA} for the control sample. Despite having a higher velocity, the observed winds in the AGN PSBs are still not powerful enough to sweep significant amounts of gas out of the halos of the host galaxies. We also detect winds of similar velocities in the stacked spectra of shocked and quenched PSBs. 
\end{abstract}

\begin{keywords}
galaxies: active -- galaxies: nuclei -- galaxies: Seyfert -- galaxies: starburst -- galaxies: evolution -- galaxies: absorption lines
\end{keywords}
\section{Introduction}

Quenched post-starburst galaxies, also known as K+A galaxies \citep[e.g.,][]{Dressler+83,Goto07}, have weak or no on-going star-formation but have unusually large A-star populations, that are indicative of a recently terminated starburst. These galaxies may be in transition from the blue cloud populated by young and star-forming galaxies to the red sequence inhabited by old and quiescent galaxies, rapid enough to transition within one billion years. Recent works \citep[e.g.,][]{Wild+09,Yesuf+14}, having identified the precursors of these galaxies with AGN or star-formation activity, have shown that AGNs are more common in these objects but that there is a significant time lag between the starburst and the AGN phase. This lag suggests that AGNs do not primarily quench starbursts. Although PSBs are less than $1\%-5\%$ of the total galaxy population in current and early universe, their number densities combined with their rapid evolution timescale hint that PSBs are important channels for the formation of red-sequence galaxies \citep{Wong+12,Yesuf+14,Wild+16}.

Theoretically, post-starburst (PSB) galaxies might be the aftermath of major mergers in galaxies \citep{Bekki+05,Hopkins+06,Snyder+11}. In gas-rich model mergers, tidal forces channel gas to galaxy centers and power intense nuclear starbursts and obscured AGN activity \citep{Barnes+91,DiMatteo+05,Hopkins+06}. After gas has been consumed by the starburst itself and/or expelled by stellar feedback (e.g., radiation from massive stars or supernova explosions) the remnant gas and dust obscuring the AGN may be cleared out by AGN feedback \citep{Hopkins+06,Hopkins+08,Snyder+11,Hayward+14,Ishibashi+16}.

Observationally, neutral gas outflows in low redshift $z \sim 0.1$ AGNs and starbursts have been extensively studied using \ion{Na}{i} 5889.95, 5895.92\,{\AA} doublet \citep[e.g.,][]{Rupke+05c,Martin+06,Krug+10,Sarzi+16}. There exist, however, only a few studies of winds in post-starburst galaxies, and the samples are small \citep{Sato+09,Coil+11,Tripp+11,Baron+17}. Prior to this paper, the nature of AGN-driven winds in large samples of post-starbursts has not been explored. In the rest of this section, we briefly review works relevant to galactic winds detected in absorption.

\citet{Krug+10} studied outflows in 35 infrared-faint (i.e., low star-forming) Seyferts in an effort to separate the starburst effects on the winds from the AGN effects. The authors compared the outflow properties of the infrared-faint Seyferts with those of infrared-bright composite Seyferts in which both starbursts and AGN co-exist. The wind velocities of both high and low starforming Seyferts are similar to those of starburst galaxies. The average wind velocity for the infrared-faint Seyferts 2 galaxies is $v_w =-137\pm  8$ km s$^{-1}$. Likewise, \citet{Rupke+05c} studied a sample of 26 Seyfert ultra-liminous infrared galaxies (ULIRGs). They found no significant differences between the velocities of Seyfert 2s, which are ULIRGs ($v_w =-456^{+330}_{-191}$\,km\,s$^{-1}$), and starbursts of comparable infrared luminosities ($v_w =-408^{+224}_{-191}$\,km\,s$^{-1}$). At $z \sim 1$, \citet{Yesuf+17a} also did not find significant wind velocity differences between X-ray detected low-luminosity AGNs and X-ray non-detected star-forming galaxies \citep[see also][]{Coil+11,Cimatti+13}.
 
Similarly, \citet{Sato+09} found \ion{Na}{i} outflow speeds of $\sim 100$\,km\,s$^{-1}$ in $\sim 10$ fading post-starburst galaxies with low-level nuclear activity at $0.1 < z < 0.5$. Within a similar redshift range,  \citet{Coil+11} also found low velocity winds ($\sim -200$\,km\,s$^{-1}$) in 13 post-starbursts galaxies by using near UV \ion{Mg}{ii} and \ion{Fe}{ii} absorption-lines. In contrast, \citet{Tremonti+07} observed high-velocity winds (with median  \ion{Mg}{ii} velocity of $v_w \sim -1100$\,km\,s$^{-1}$) in massive transitional post-starburst galaxies. In subsequent works, they argued that these fast outflows are most likely driven by feedback from extremely-compact, obscured starbursts rather than AGN \citep{Diamond-stanic+12,Geach+14,Sell+14}. But, it remains possible that the outflows were driven by AGN activity that has been recently switched off. \citet{Sell+14} found low-luminosity AGN in half of their post-starburst sample.

\citet{Alatalo+16a} studied the strength of \ion{Na}{i} absorption in PSBs which show shock emission-line ratios (hereafter referred to as shocked PSBs). They found that these galaxies have significant excess \ion{Na}{i} absorption relative to \ion{Mg}{i} b compared to other emission-line galaxies. The authors speculated that the \ion{Na}{i} enhancement may be related to AGN-driven winds. However, they did not analyze the \ion{Na}{i} profile with a wind model. 

In this work, using a simple wind model, we analyze in detail the \ion{Na}{i} profile of post-starbursts, including those that show AGN or shock signatures in their spectra. We use high signal-to-noise stacked spectra of a large sample of post-starburst galaxies and carefully selected control samples from the Sloan Digital Sky Survey (SDSS). We confirm that the shocked PSBs have excess \ion{Na}{i} absorption but find that only a small fraction of the absorption is due to wind absorption.

\citet{Chen+10} have done similar analyses using all SDSS galaxies. They, however, excluded AGN in their sample and have not focused on studying winds in post-starburst galaxies. Nevertheless, for normal SDSS galaxies, they found that the \ion{Na}{i} absorption strength depends strongly on dust attenuation (A${\rm_V}$) and star-formation surface density. But, the mean outflow velocity does not depend strongly on any galaxy physical parameters in the limited dynamic range probed by their study. The strength of \ion{Na}{i} absorption near the systemic velocity also correlates with inclination angles of galactic disks. Note also that scattered resonant emission infill of the absorption-line may affect edge-on systems \citep{Prochaska+11}. \citet{Chen+10} found emission components on top of \ion{Na}{i} absorptions in edge-on galaxies. We use the inclination angle, dust attenuation, and other physical quantities as control parameters to isolate the effect of AGN on the wind velocity and strength.

This paper is different from preceding works for the following reasons : 1) It presents the largest stacked \ion{Na}{i} profile analysis for local PSBs \citep[cf.][]{Sato+09,Krug+10,Coil+11} 2). It has better matched control samples, and some of the previous works did not have control samples at all \citep[cf.][]{Sato+09,Coil+11}. 3) It is focused on PSBs or AGN PSBs. Our analysis and method are similar to \citet{Chen+10} but these authors excluded AGNs, and studied mainly non-post-starburst galaxies. Very recently, \citet{Nedelchev+17}  studied $\sim 9900$ Seyfert 2 galaxies (most are not PSBs) in SDSS but only 53 Seyferts show winds in \ion{Na}{i} absorption. We, however, detect winds in stacked spectra AGN (Seyfert) PSBs, implying that the winds are ubiquitous. A detailed comparison of our work with \citet{Nedelchev+17} will be a subject future paper.

This paper is organized as follows: section~\ref{sec:samp} presents the sample selection of post-starbursts and their control sample. In section~\ref{sec:analy}, we present our results and analyses of how we stack the spectra and fit a stellar population synthesis model and a simple wind model to the stacked spectra. Section~\ref{sec:disc} discusses the  \ion{H}{i} mass implied by the inter-stellar medium (ISM) component of \ion{Na}{i} profile, the wind mass outflow rate inferred from the wind component and the  ionized wind velocity and outflow rate detected using \ion{O}{iii} emission in the stacked spectrum of AGN PSBs. Section~\ref{sec:conc} presents the main conclusions of the paper. A wavelength measured in air (not vacuum wavelength) is given throughout the paper.  For notational convenience, we express the percentiles as $\pm$ deviations from the median from now on. For instance, $X^{+Y}_{-Z}$ denotes that X is the median, $X+Y$ is the 84th percentile and $X-Z$ is the 16th percentile. For a Gaussian density function, Y and Z equal to its standard deviation.\\
 
\section{Sample Selection}\label{sec:samp}

\subsection{Post-starburst Selection}

The data for post-starburst and control samples come from the Sloan Digital Sky Survey \citep[SDSS,][]{Alam+15}. We use the SDSS measurements for physical quantities such as stellar masses and spectral indices in the publicly available Catalog Archive Server (CAS) \footnote{http://skyserver.sdss.org/casjobs/}. 

\begin{figure}
\includegraphics[width=0.9\linewidth]{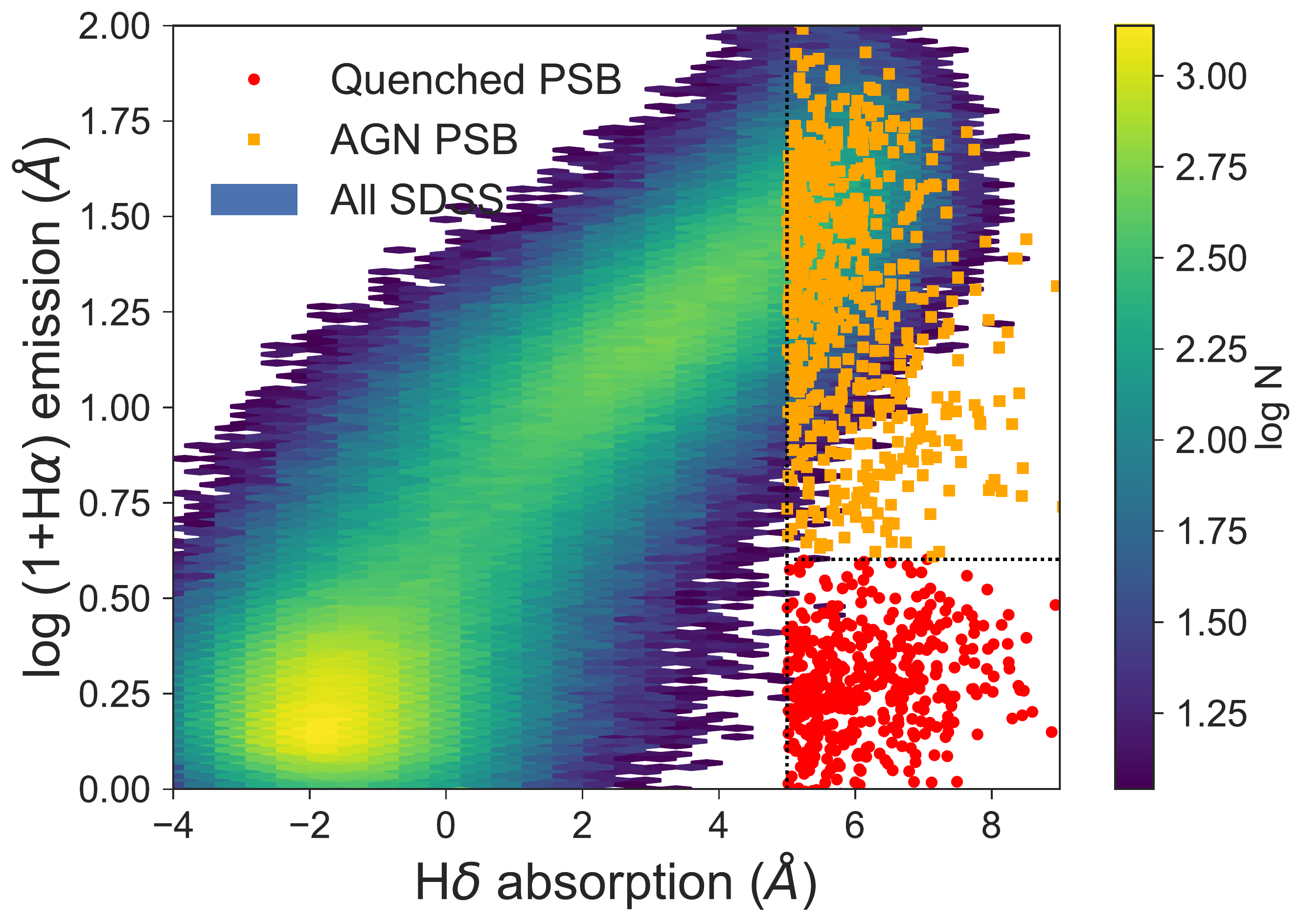}
\caption{H$\delta$ absorption equivalent width versus H$\alpha$ emission equivalent width. The colorbar show the number density of the SDSS galaxies above log M (M$_\odot$) > 10 at $z=0.05-0.25$. The red points at the bottom right corner are recently quenched PSBs ($N=447$), defined using the cuts denoted by the broken lines and with the additional cut of \ion{O}{ii} < 3{\AA} \citep[cf.][]{Goto07}. The orange squares are active galactic nuclei post-starbursts (AGN PSBs, $N=560$), which are identified as AGN using the BPT emission-line ratio diagnostic. \label{fig:hd_ha}}
\end{figure}

Following \citet{Goto07}, we define quenched PSB galaxies as objects with H$\delta$ absorption above 5{\AA}, and \ion{O}{ii} and H$\alpha$ emission below 3{\AA}. Similarly, we define AGN PSBs as objects with  H$\delta$ absorption above 5{\AA} that are classified as AGN based on the BPT line ratio AGN diagnostic \citep{Baldwin+81,Kewley+01,kauffmann03c}. Specifically, we use the \emph{bptclass=4} flag in the \emph{galSpecExtra}\footnote{http://www.sdss.org/dr12/spectro/galaxy\_mpajhu/} catalog, which identifies well-measured, strong, AGN (excluding low ionization emission-line galaxies, LINERs). In both quenched and AGN PSB selection, we require the H$\delta$ absorption be measured above 2$\sigma$, the median signal to noise of the spectra be at least ten and the redshift range be $z =0.05-0.25$. We exclude broad-line AGNs by requiring the velocity dispersion measured simultaneously in all of the Balmer lines to be below 200 \kms. The presence of a broad-line AGN complicates the stellar population modeling, and existing models, including the one use in this paper, do not properly model galaxies that host broad-line AGNs. Figure~\ref{fig:hd_ha} shows the two post-starburst samples on H$\delta$ absorption versus H$\alpha$ emission equivalent width space. PSBs deviate from the locus of normal galaxies when their star-formation quenches rapidly. The stacked spectrum of quenched PSBs is shown in Figure~\ref{fig:qpsb_spec}.

With the above criteria, we select 447 quenched PSBs and 560 AGN PSBs. We adopted a simple, albeit incomplete, post-starburst AGN selection compared to that of \citet{Yesuf+14} because our analysis is based on stacked spectra, and the selected samples will likely characterize well the average properties of the samples. This simplified approach was also adopted in \citet{Alatalo+16a}, which motivated this follow-up work. We show in the Appendix that restricting the AGN PSB sample to H$\delta$ above 6{\AA}, in order to increase the purity of the sample, does not change our conclusions. We also stack the spectra of 1060 shocked PSBs identified by \citet{Alatalo+16a} and analyze their \ion{Na}{i} profile to study their wind properties.

\subsection{Control Sample Selection}\label{sec:cont}

Star-formation rates (SFRs), viewing angles and dust properties of galaxies affect the profile shapes of the \ion{Na}{i} doublet \citep{Chen+10}. We control for these effects and isolate the AGN effect for a given target AGN PSB (or shocked PSB), by selecting  its counterpart using the criteria listed below. 

A control galaxy is :
\begin{enumerate}[label={\arabic*.}, noitemsep,leftmargin=\parindent]
\item purely star-forming and does not have emission-line ratios indicative of AGN 
\item within stellar mass $\abs{\Delta \, \log M (M_\odot)} < 0.1$ of the target
\item within redshift $\abs{\Delta \, z} < 0.03$  of the target PSB
\item within axis-ratio $\abs{\Delta \, b/a} < 0.1$ of the target PSB
\item within bulge fraction $\abs{\Delta \,fracDev} < 0.3$  of the target
\item within 4000\,{\AA} break $\abs{\Delta\, D_n(4000)}  < 0.2$ of the target
\item within Balmer decrement $\abs{\Delta\, H\alpha/H\beta}  < 1$ of the target
\item within WISE 12\,$\mu$m to 4.6\,$\mu$m flux ratio $\abs{\Delta \, f_{12}/f_{4.6}}  < 0.2$ of the target PSB
\end{enumerate}

Accurately estimating the SFR of  an AGN-host galaxy is challenging since the H$\alpha$ is contaminated by the AGN. $D_n(4000)$ and $f_{12}/f_{4.6}$ are crude star-formation rate indicators on longer time scales that also work for AGNs \citep{Brinchmann+04,Donoso+12}. For the dust-free Case B recombination, the H$\alpha$/H$\beta$ ratio is 2.86 for star-forming galaxies and 3.1 for AGNs. Our seventh matching criterion, $\abs{\Delta\, H\alpha/H\beta}  < 1$, is crude to accommodate such subtle differences between AGN PSBs and their control galaxies. Since the SDSS fiber does not entirely cover a galaxy, the H$\alpha$/H$\beta$ ratio reflects the central dust attenuation. To assess the global dust attenuation, we adopt global $f_{12}/f_{4.6}$ WISE ratio, although
the dust heating mechanisms in PSBs are not well understood.

\begin{figure*}
\centering
\includegraphics[width=0.4\linewidth]{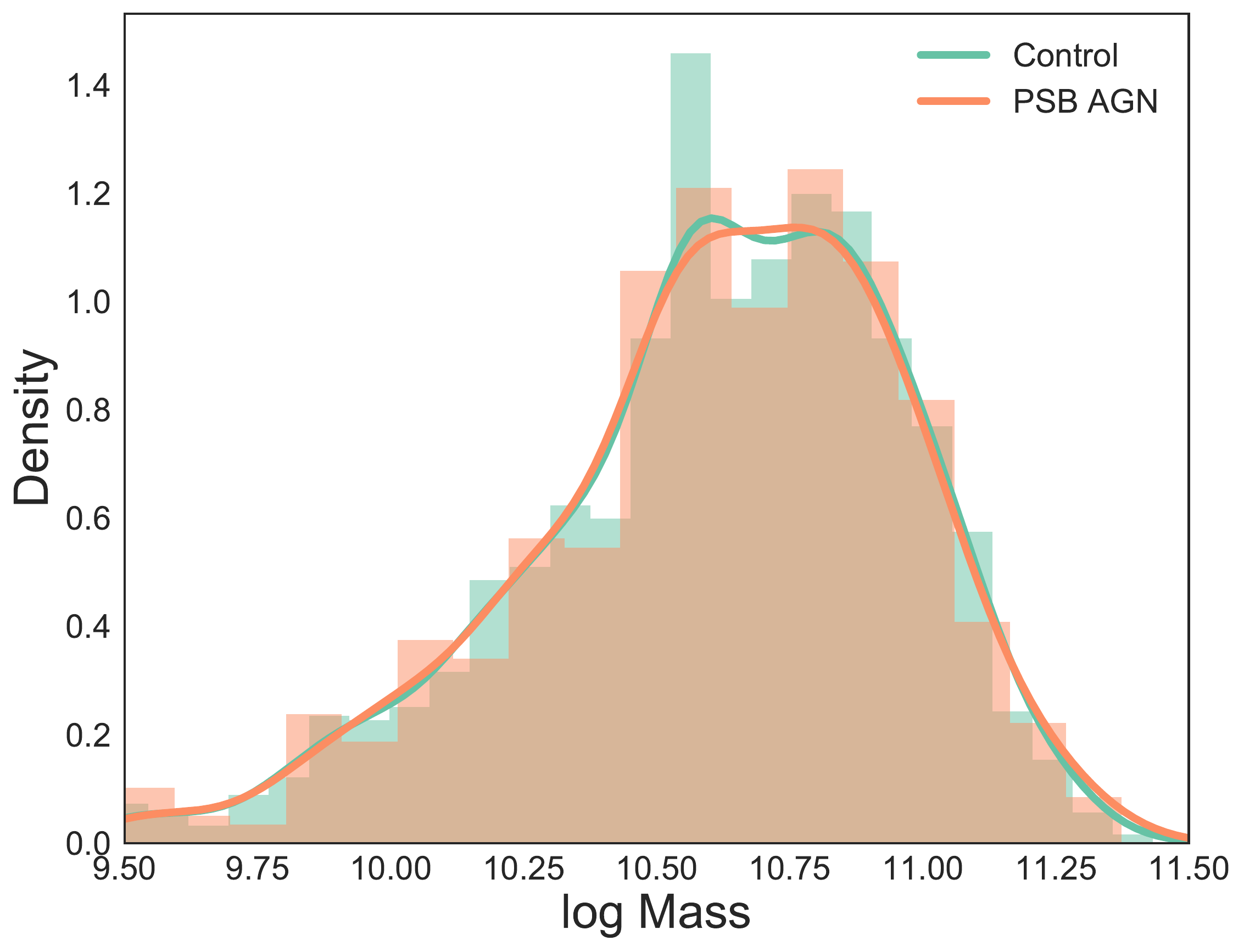}
\includegraphics[width=0.4\linewidth]{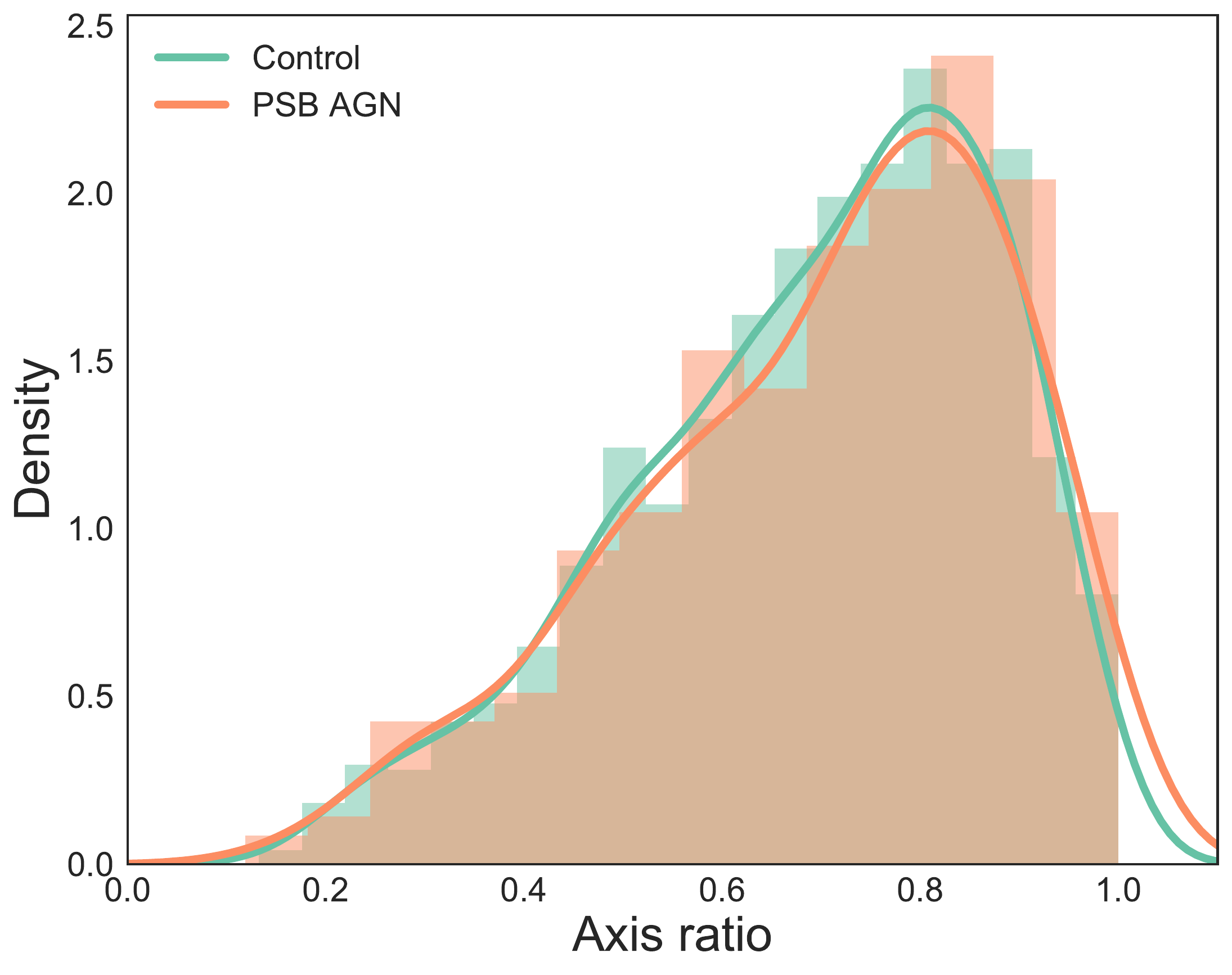}
\includegraphics[width=0.4\linewidth]{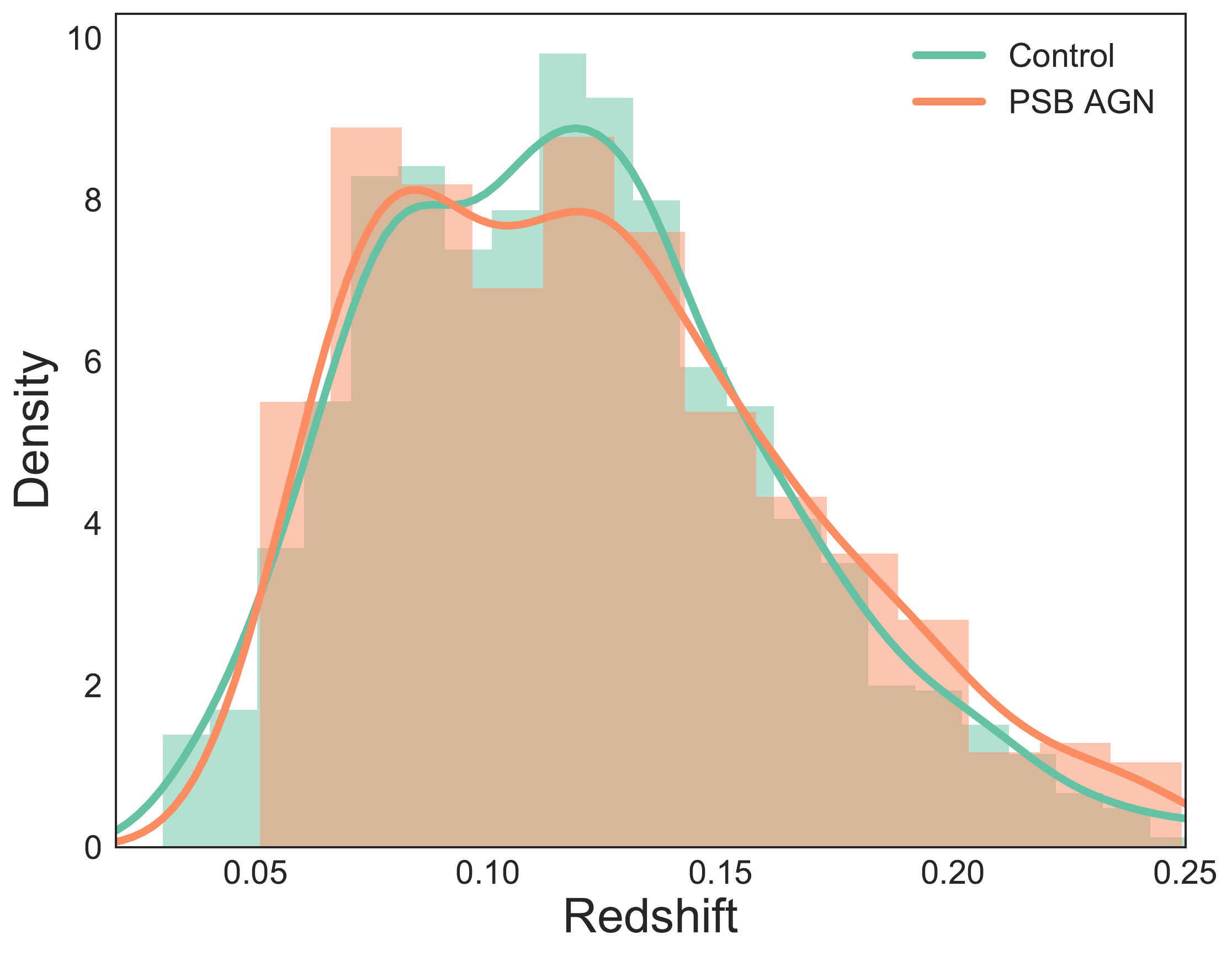}
\includegraphics[width=0.4\linewidth]{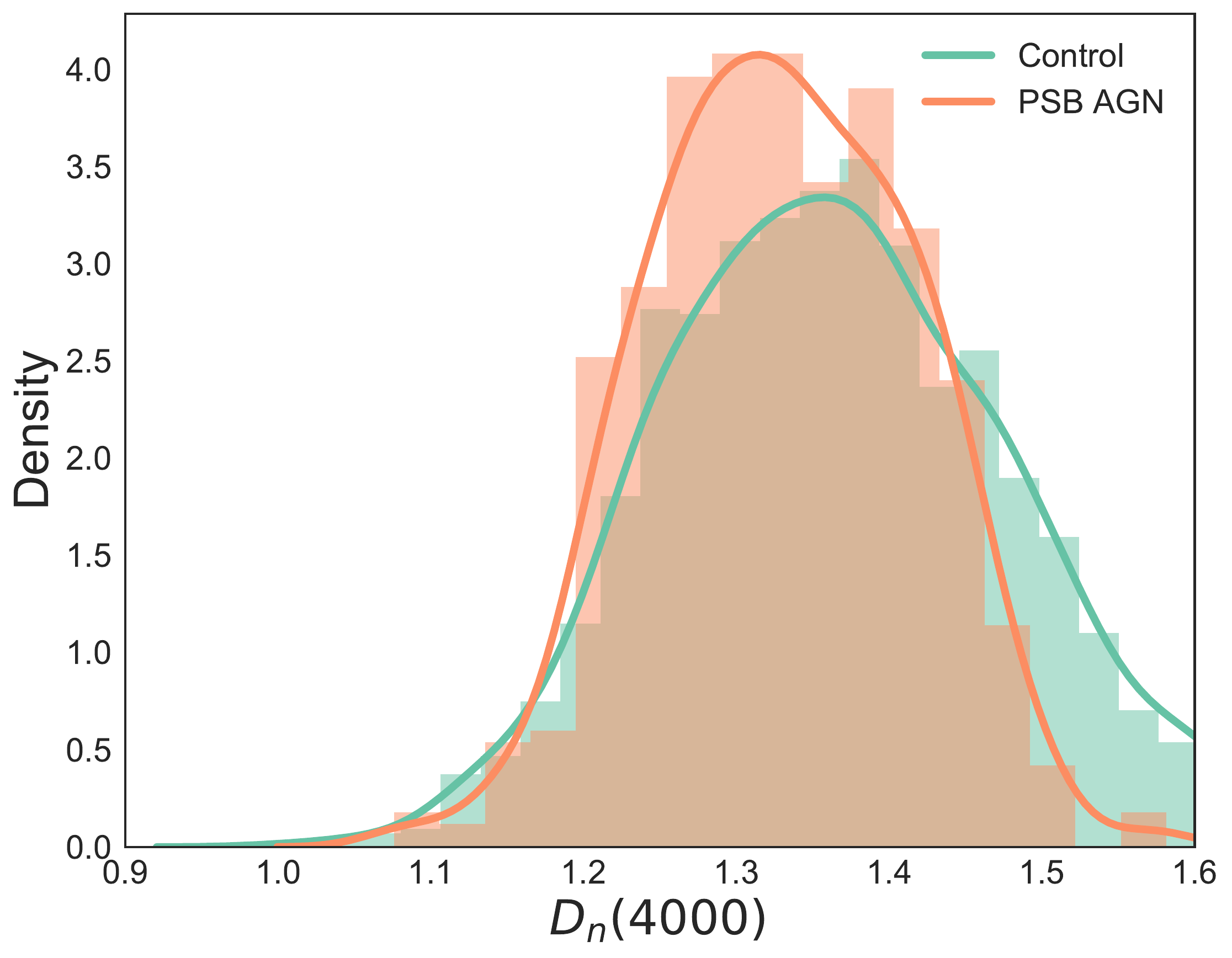}
\includegraphics[width=0.4\linewidth]{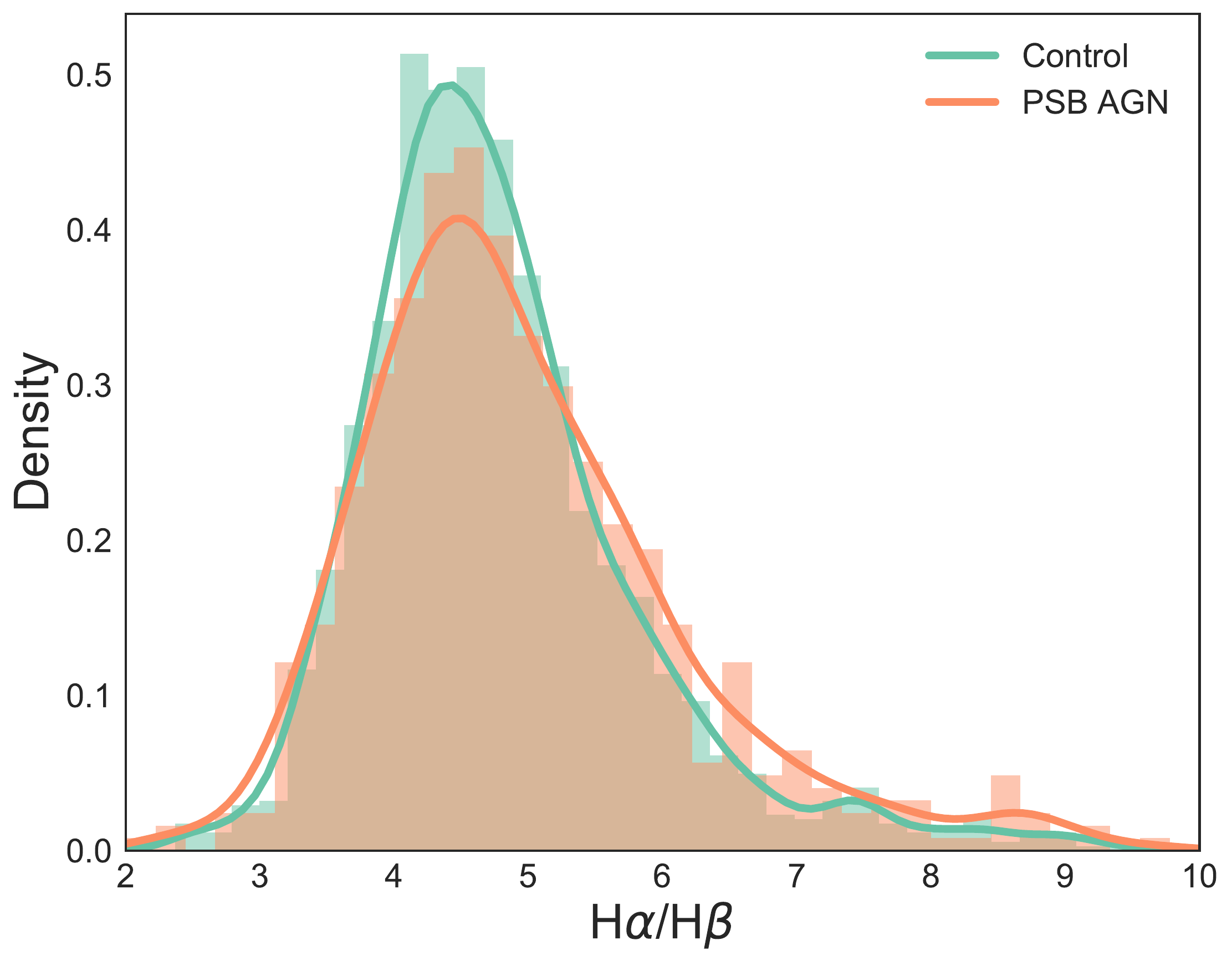}
\includegraphics[width=0.4\linewidth]{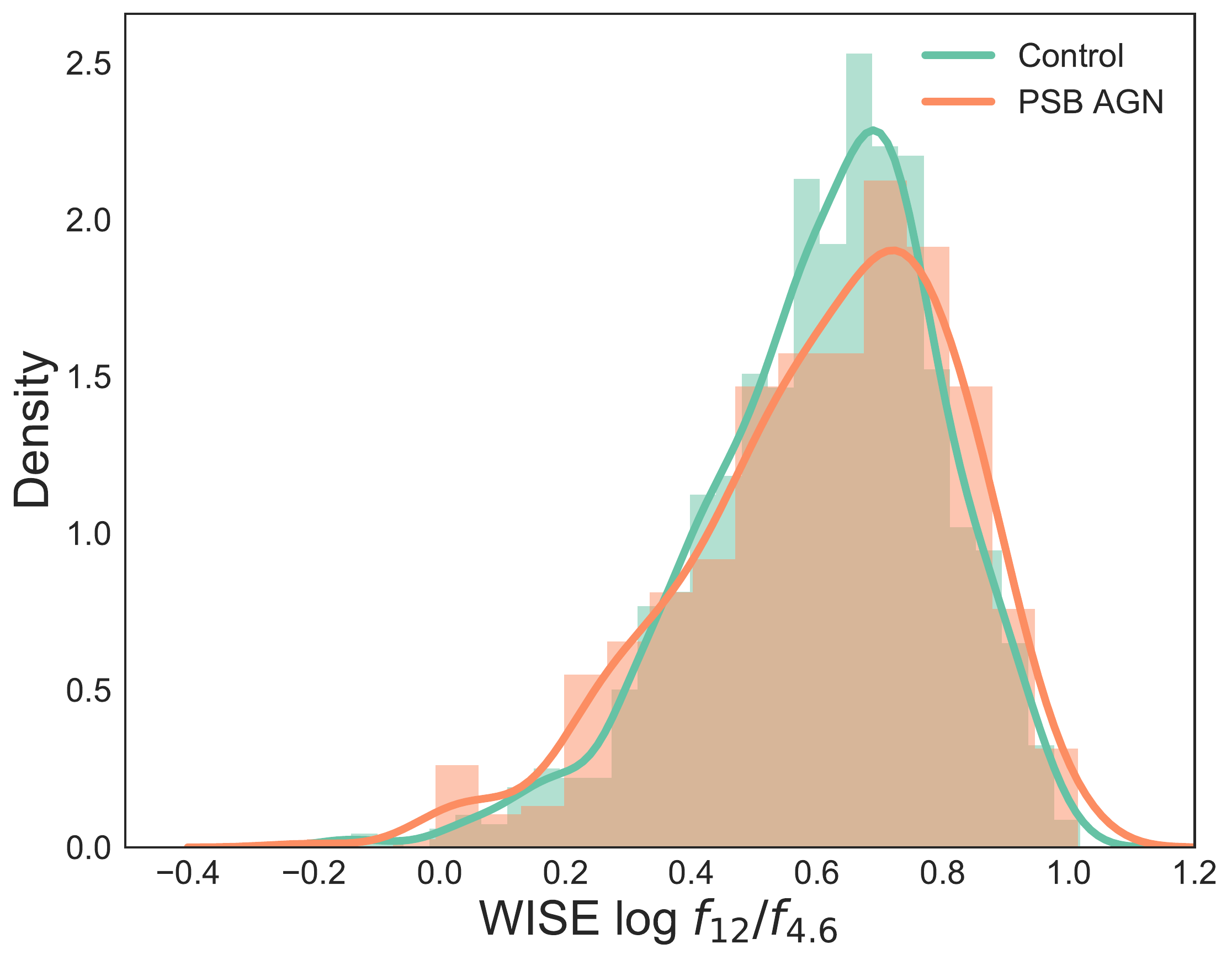}
\caption{The density distributions for the control sample selection parameters for AGN PSBs and their control sample. In an attempt to isolate the AGN effect and account for SFR, viewing angle, and dust absorption effects, a control sample is constructed by matching in stellar mass, axis-ratio, redshift, 4000\,{\AA} break, H$\alpha$ to H$\beta$ flux ratio, and WISE 12\,$\mu$m to 4.6\,$\mu$m flux ratio. The $D_n(4000)$ does not much well but the scatter in $D_n(4000)$ versus specific star-formation relation is larger than difference \citep{Brinchmann+04}. The next figure, will show that the H$\alpha$ emissions in the AGN PSBs and the control sample are on average comparable, implying that the instantaneous SFRs of AGN PSBs are similar to that the control sample or even lower. Note that the control galaxies are not necessarily PSBs without AGNs. The result of such control sample are discussed in the appendix. \label{fig:matchdist}}
\end{figure*}

About 98\% of the AGN PSBs have at least one control match that satisfies the above criteria. If an AGN PSB has more than three matches, we randomly select only three of them. About 95\% of AGN PSBs have three random matches that satisfy the above criteria. Figure~\ref{fig:matchdist} shows the resulting distributions for stellar mass, axis ratio, redshift, $D_n(4000)$, H$\alpha$/H$\beta$ and $f_{12}/f_{4.6}$ of the AGN PSBs and their control sample. There are reasonable agreements between the distributions of the two samples.  Similarly, about 90\% of the shocked PSBs also have at least one match that satisfies the above criteria and have distributions that are in good agreement with their control sample.


\section{Analysis \& Results}\label{sec:analy}

\subsection{Coadding Spectra}
To coadd the logarithmically-binned, observed, SDSS spectra of PSBs or their comparison samples, each spectrum is shifted to its rest frame wavelength and is normalized by the mean flux in the wavelength range 5445 -- 5550{\AA} in logarithmic space. The rest-frame spectra are then interpolated on a logarithmic wavelength grid with bin $\Delta \lambda = 10^{-4}$ dex, which is the same as the pixel spacing of the observed spectra. The observed fluxes at a given wavelength bin of the rest-frame spectra are then co-added and divided by the sample size to get the mean stacked spectrum. The standard errors of the mean spectrum are estimated using a bootstrap scheme. Namely, we repeat the aforementioned stacking procedure 500 times by resampling with replacement a subset of individual spectrum with equal size as the original sample. The standard deviation of the mean-normalized fluxes of all 500 composite spectra in a given wavelength bin is the error of the mean normalized flux at that given bin.

\begin{figure*}
\centering
\includegraphics[width=0.9\linewidth]{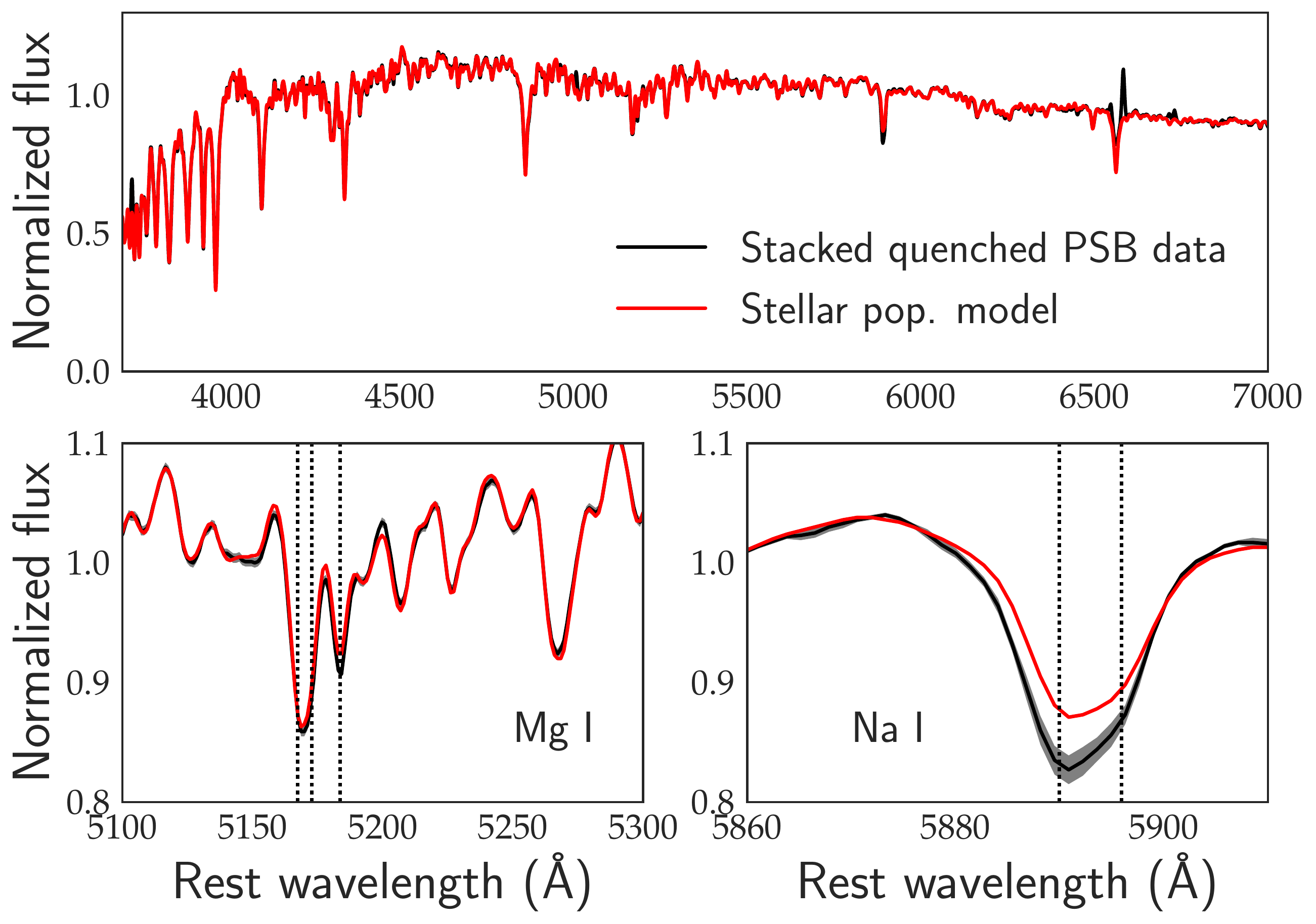}
\caption{The stacked spectrum of 447 quenched post-starbursts (black) fitted with a stellar population synthesis model (red). Quenched PSBs are characterized by strong Balmer absorption-lines such as H$\delta$ 4102. The observed stellar continuum is well fit by stellar population model but there is residual \ion{Na}{i} absorption due to interstellar and galactic wind absorption. Characterizing such \ion{Na}{i} residual in post-starburst galaxies is the subject of this work. The \ion{O}{ii} 3727 \& \ion{N}{ii} 6583 emission-lines are due to LINER like emission and are perhaps related to weak AGN in these galaxies. Note that the \ion{Na}{i} doublet is a resonant line arising from the ground state ($0.0 \rightarrow 2.1$ eV) while the \ion{Mg}{i} triplet arises from a transition between excited states ($2.7 \rightarrow 5.1$ eV). The latter is, thus, not observed in the cold neutral ISM. \label{fig:qpsb_spec}}
\end{figure*}

\subsection{Stellar Continuum Modeling}
It is well known that the \ion{Na}{i} 5889.95, 5895.92\,{\AA} doublet arises due to photospheric stellar absorption and inter-stellar medium (ISM) absorption. To account for the stellar absorption, we fit our stacked spectra using a publicly available penalized pixel fitting (PPXF) code \citep{Cappellari+04,Cappellari+17}. After masking out spectral regions potentially affected by major emission-lines, the code determines the optimal linear combinations of single stellar population model templates from \citet{Vazdekis+10}. 
The templates are based on the MILES library and have spectral resolutions comparable to the observed spectra.

In addition to the lines in the code's default line list, we have masked skyline regions 5577, 6300, and 6363{\AA} and nebular lines,  \ion{Ne}{iii} 3869, \ion{He}{ii} 4685, \ion{N}{i} 5197, \ion{He}{i} 5876, and \ion{Ar}{iii} 7136{\AA}. The \ion{Na}{i} 5889.95, 5895.92 doublet is also masked out because it has a substantial residual that was not fit by the stellar population templates due to interstellar absorption. We use 12th-order additive polynomial corrections for the spectral shape of the stellar templates. Although the continuum fit parameters are not unique and the continuum modeling is degenerate, the model, nevertheless, does a good job of reproducing the stellar absorption-lines and the continuum shape. We have tried varying the continuum fitting parameters and  learned that our main conclusions are not impacted by these variations. For the purpose of the current work, the continuum model is adequate. Figure~\ref{fig:qpsb_spec} illustrates continuum fit for the stacked spectrum of quenched post-starbursts. The stellar population model fits the data well but there is a residual \ion{Na}{i} absorption, which we will later interpret as interstellar and galactic wind absorptions. We will present the analysis of the \ion{Na}{i} profile with a wind model later.\\

\begin{figure*}
\includegraphics[width=0.9\linewidth]{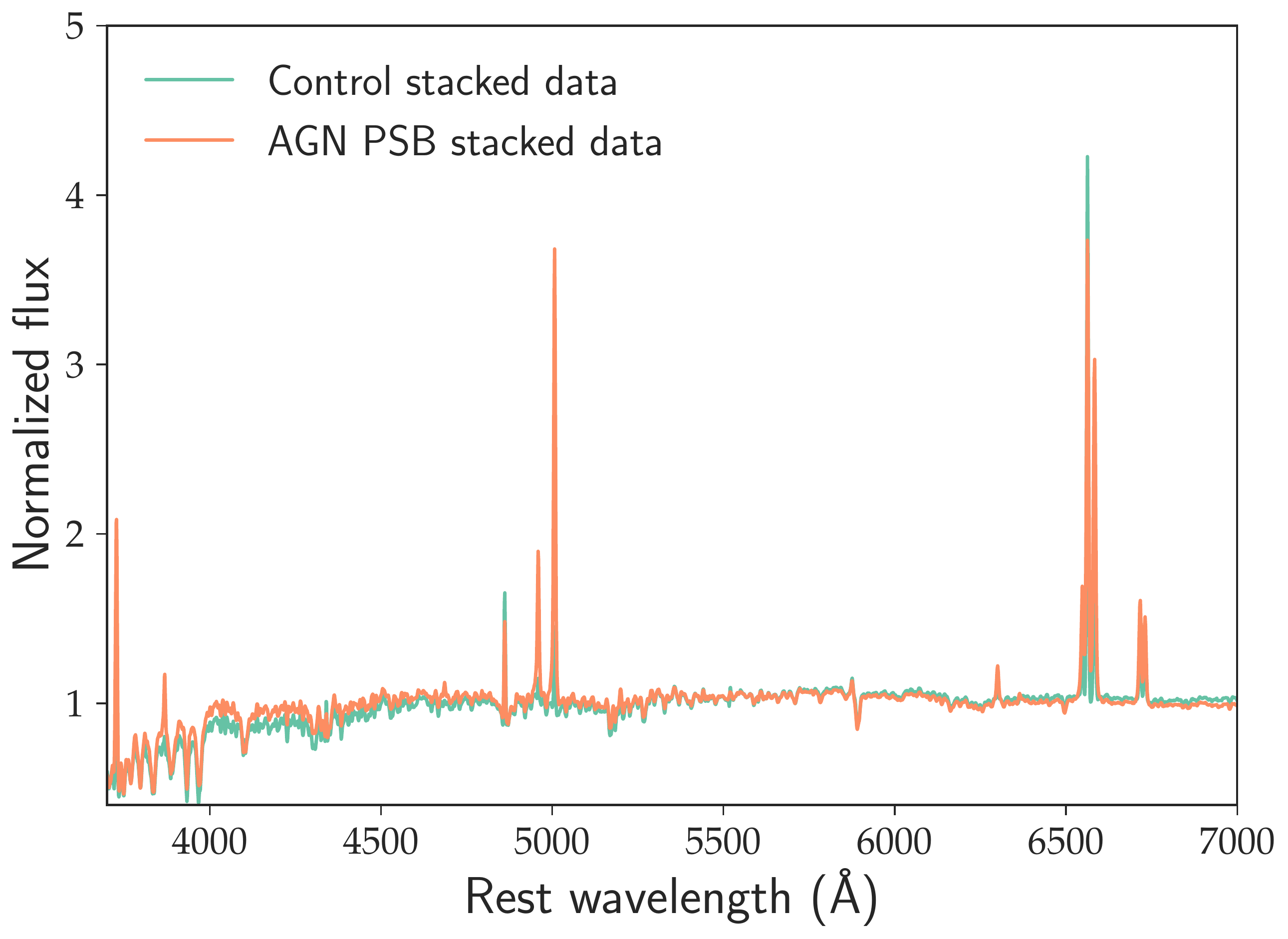}
\includegraphics[width=0.45\linewidth]{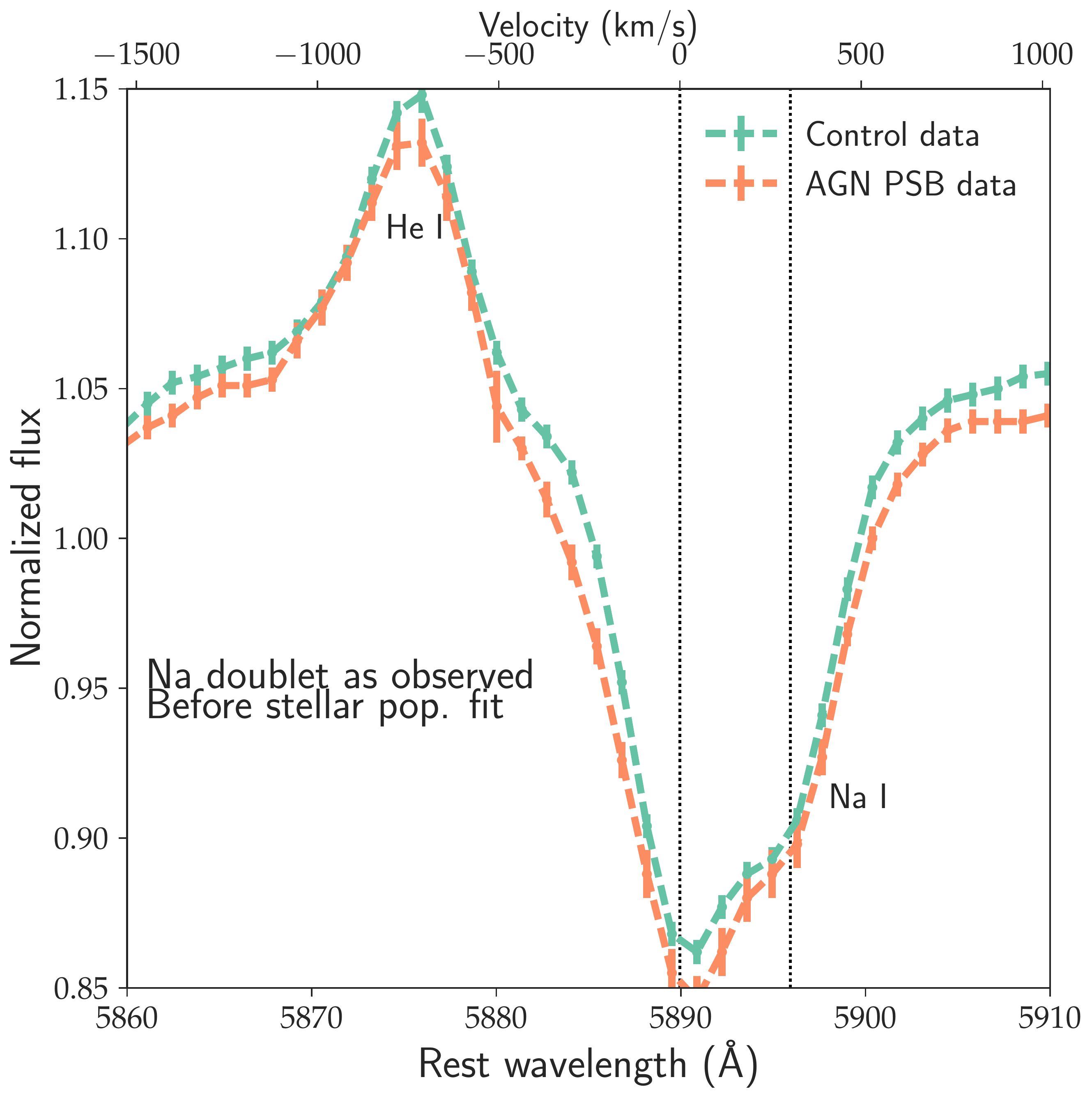}
\includegraphics[width=0.45\linewidth]{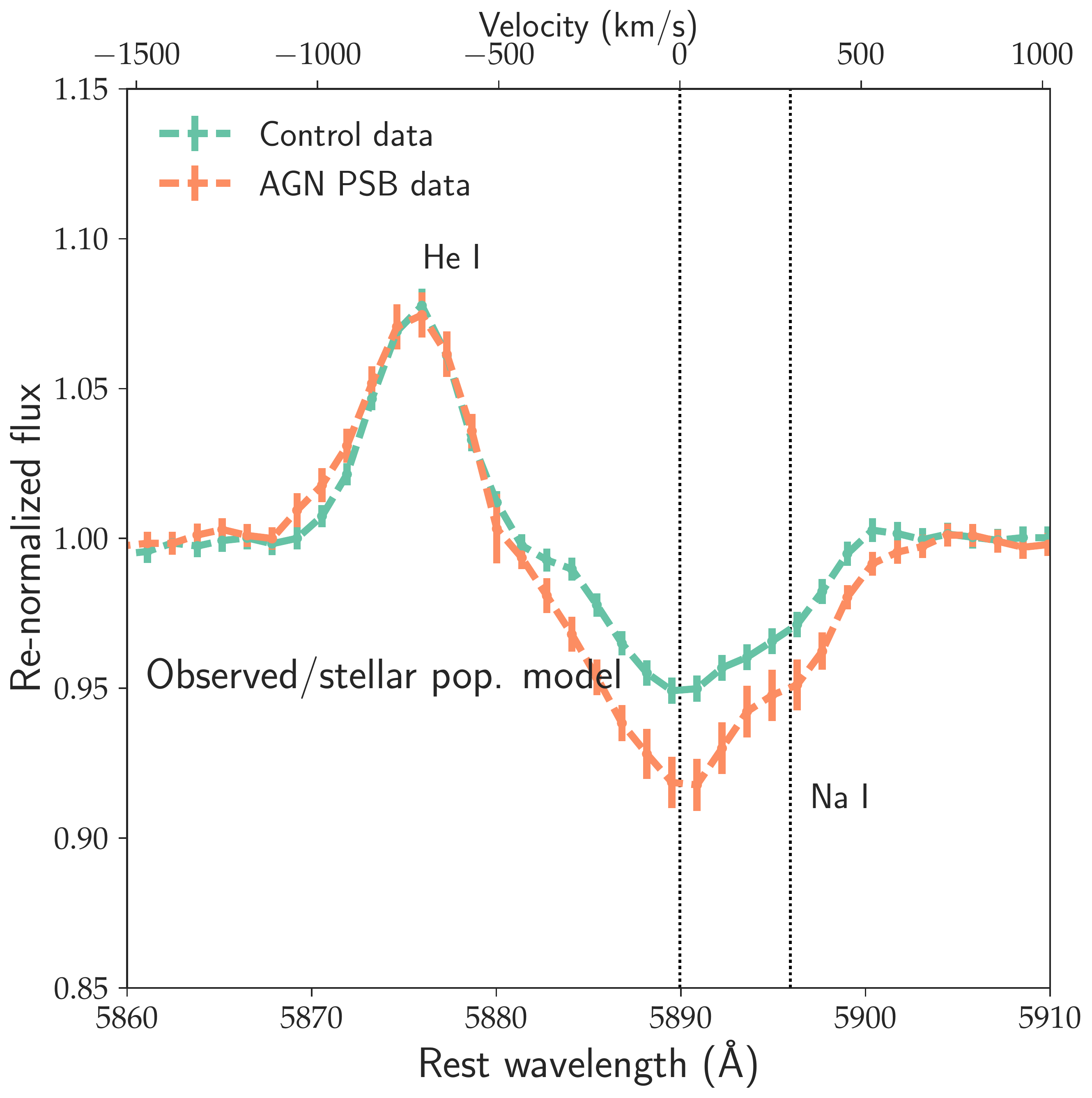}
\caption{Comparison of the stacked spectra of post-starburst AGNs with a well-matched control sample in quantities shown in Figure~\ref{fig:matchdist}. Overall, the continua of the two samples agree well but, as expected, the emission-line ratios are different because of AGN activity or lack thereof. From the strength of the H$\alpha$ emission-line, one can infer that AGN PSB have lower star-formation rates than the control sample. So, any excess in wind velocity in these galaxies must be associated with AGN. 
The continuum around 4000\,{\AA} is higher in AGN PSBs than in the control sample likely due to the larger fraction of intermediate age stars following the recent starburst. The bottom panels show zoomed-in portion around \ion{Na}{i} absorption before and after re-normalization by the stellar population synthesis model. The \ion{Na}{i} absorption profiles of the AGN and of the control sample are nearly identical. An excess, high-velocity-tail is apparent in the AGN profile and it will be quantified in the next figures.\label{fig:agn_cont_spec}}
\end{figure*}

\begin{figure*}
\centering
\includegraphics[width=0.8\linewidth]{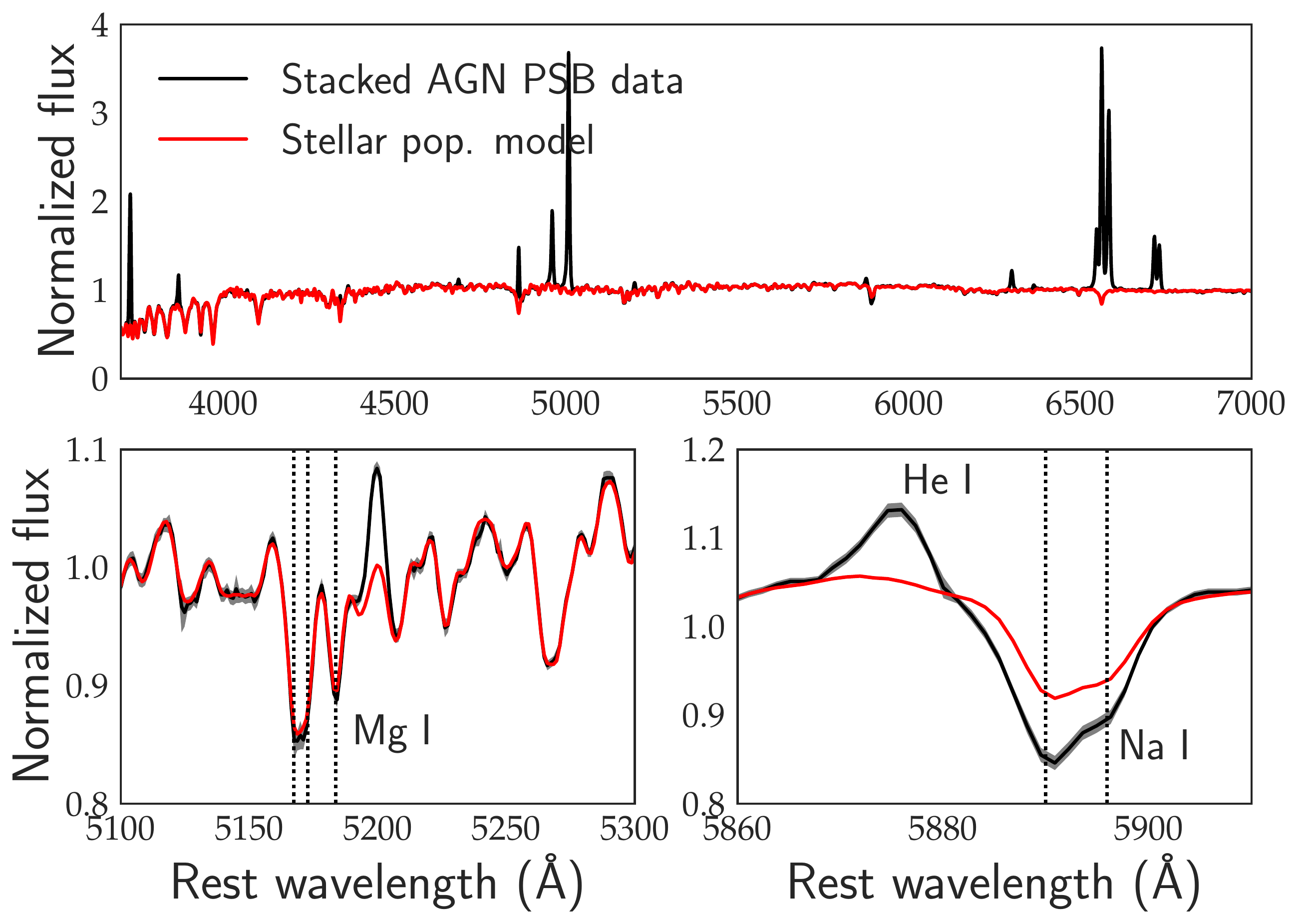}
\includegraphics[width=0.8\linewidth]{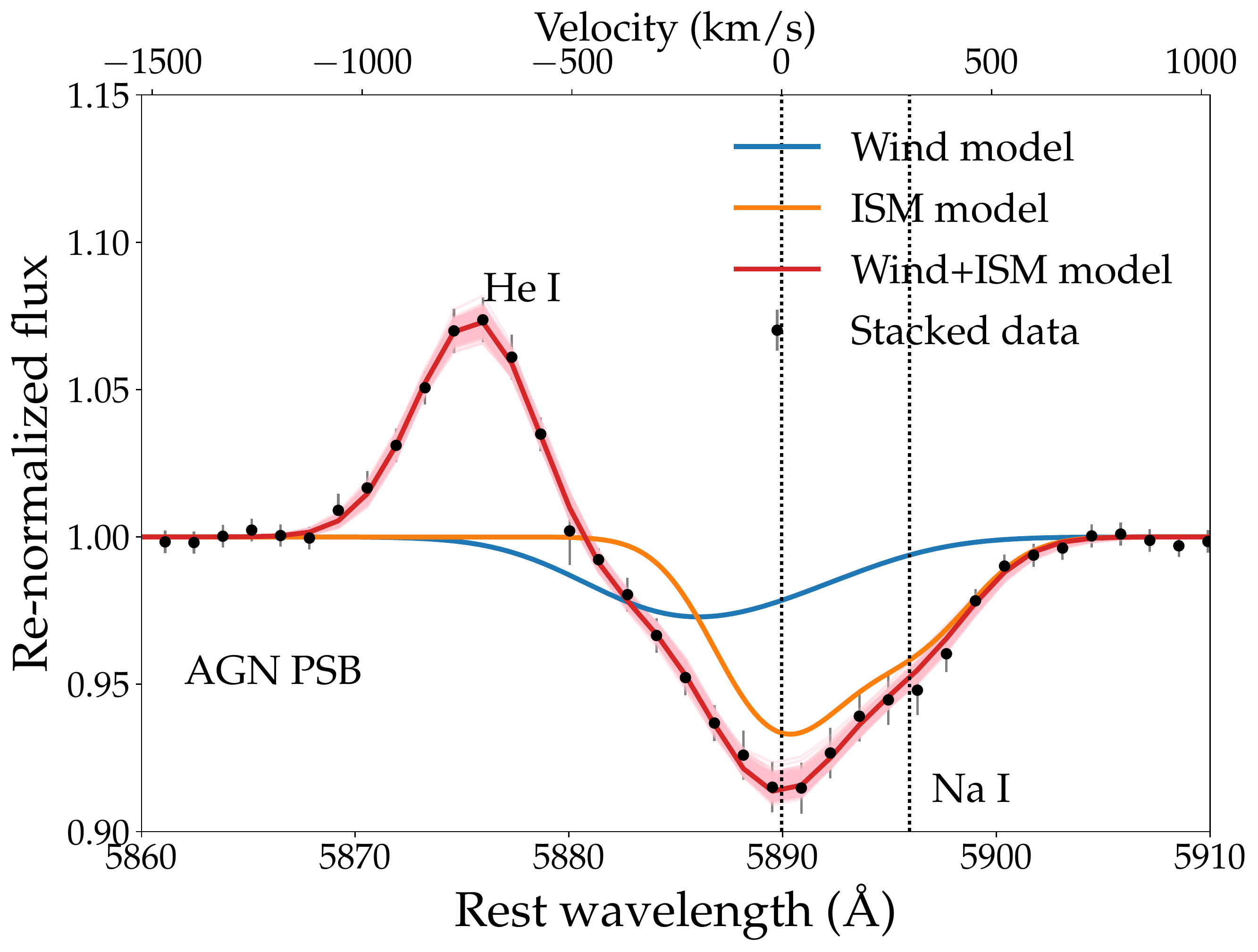}
\caption{The stacked spectrum of post-starburst AGN are fitted with a stellar population synthesis model and are continuum re-normalized by the fit. Then, the re-normalized spectra around \ion{Na}{i} are fit by a two component ISM+Wind model. There is excess \ion{Na}{i} absorption unaccounted for by stellar photospheric absorption. In the bottom panel, the data are shown as black dots with their 2$\sigma$ bootstap error bars. The red curve shows the median two component model and the pink shading denotes bands spanned by 500 random draws from the posterior probability density functions (PDFs) of the model. The blue and orange curves represent the median wind and ISM component respectively before the model is convolved with instrumental resolution. 
We infer a wind centroid velocity of $-252^{+64}_{-57}$\,\kms and maximum velocity of $-678^{+54}_{53}$\,\kms.  The equivalent width of the ISM component is $0.65^{+0.07}_{-0.09}$ {\AA} while that of the wind component is $0.36^{+0.10}_{-0.07}$ {\AA}. Note that the \ion{He}{i} 5876 nebular emission is fit together with the absorption-line. Details of the wind model and its other parameter estimates can be found in the text and Table~\ref{tab:fit}. \label{fig:apsb_fit}}
\end{figure*}

\begin{figure*}
\centering
\includegraphics[width=0.8\linewidth]{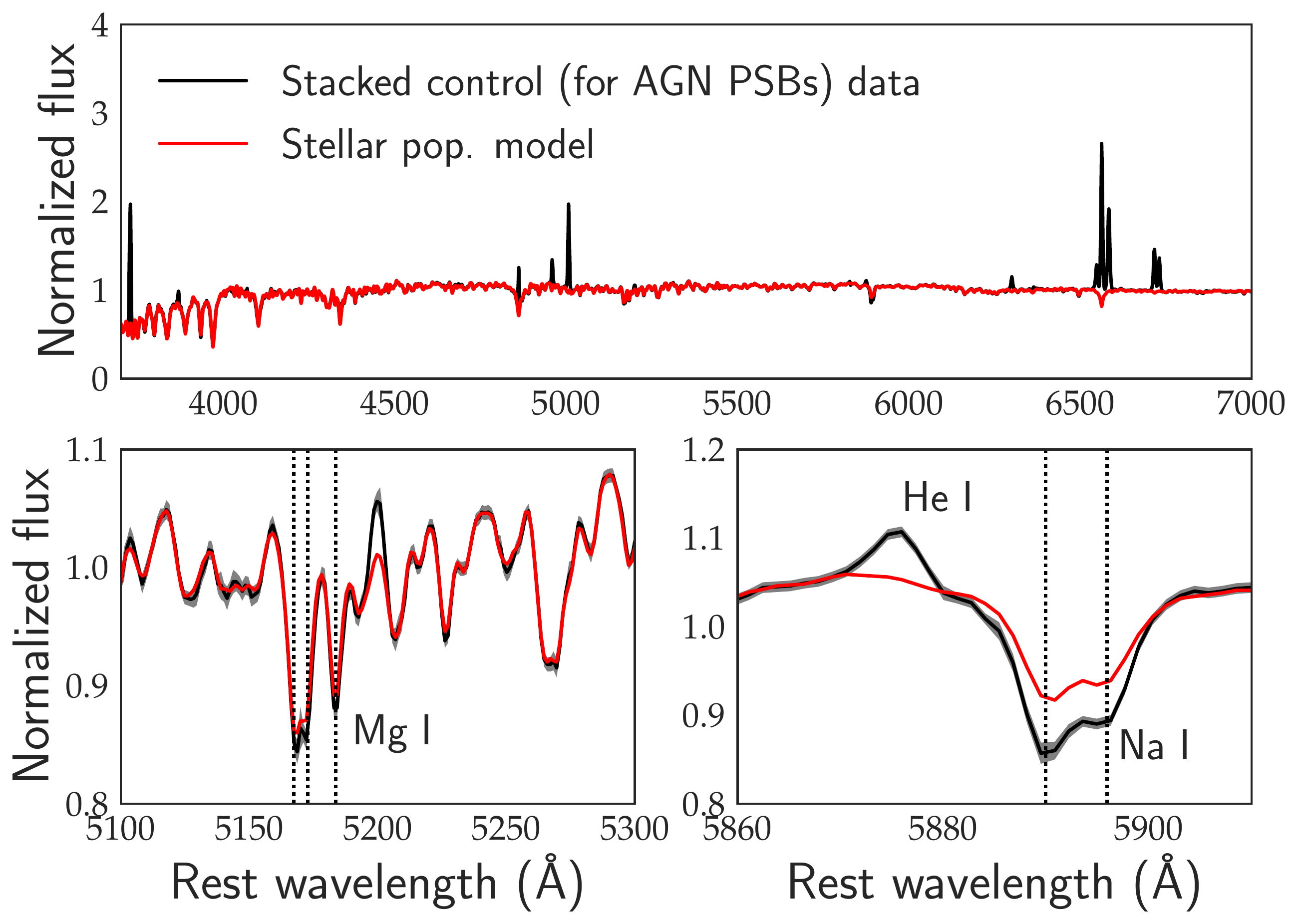}
\includegraphics[width=0.8\linewidth]{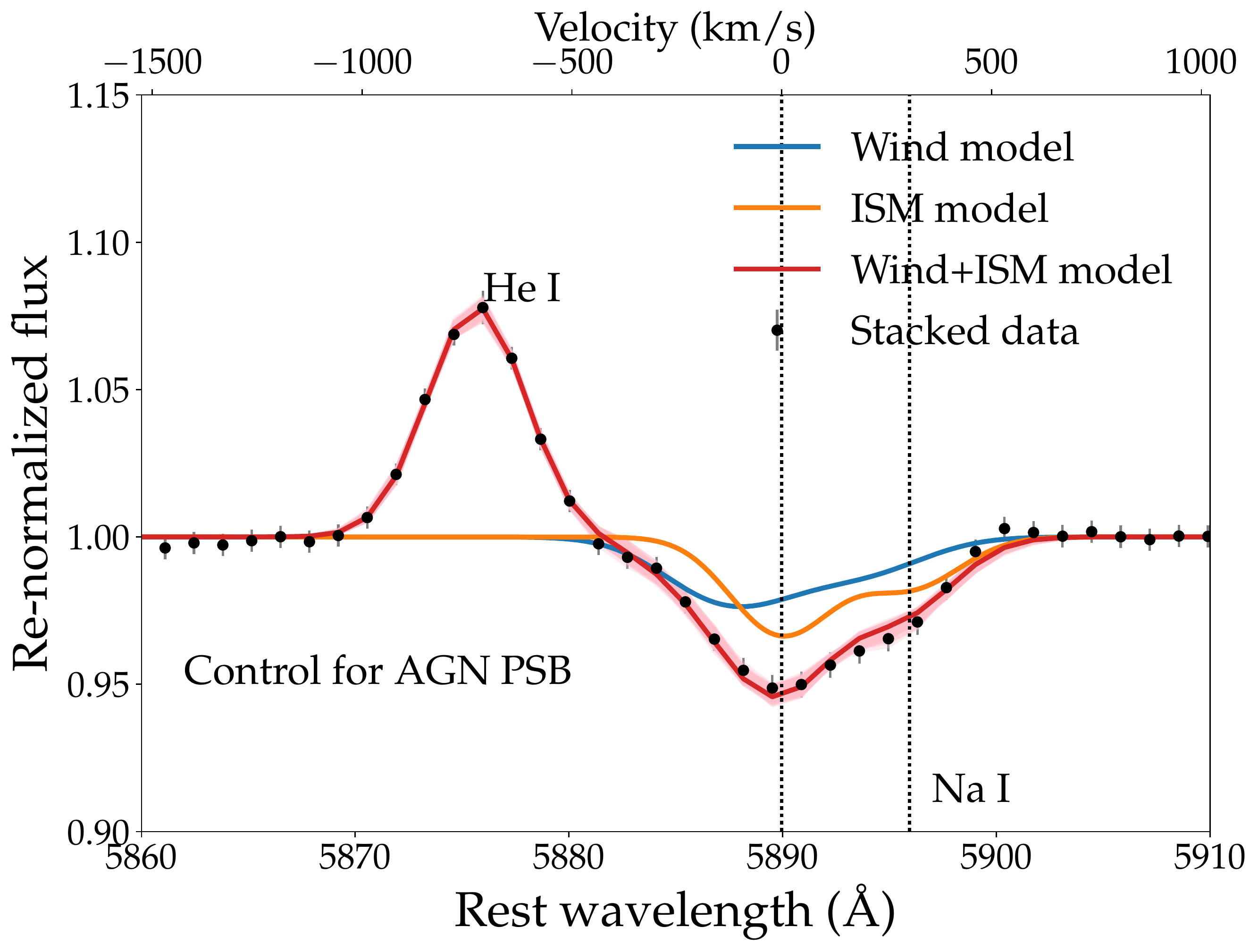}
\caption{Similar to Figure~\ref{fig:agn_cont_spec}, but here we stacked spectrum of the control sample of post-starbursts analyzed with a wind model after re-normalizing it with a stellar population synthesis model. The inferred centroid and maximum velocities are $-119^{+33}_{-41} $and $-406^{+51}_{-61}$\,\kms respectively. The equivalent width of the ISM component is  $0.30^{+0.06}_{-0.08}$\,{\AA} while that of the wind component is  $0.24^{+0.07}_{-0.06}$\,{\AA}. The centroid wind velocity and maximum velocity in the stacked spectrum of AGN PSBs are significantly higher than those in the control sample. \label{fig:apsb_cont_fit}}
\end{figure*}

\begin{figure*}
\centering
\includegraphics[width=0.8\linewidth]{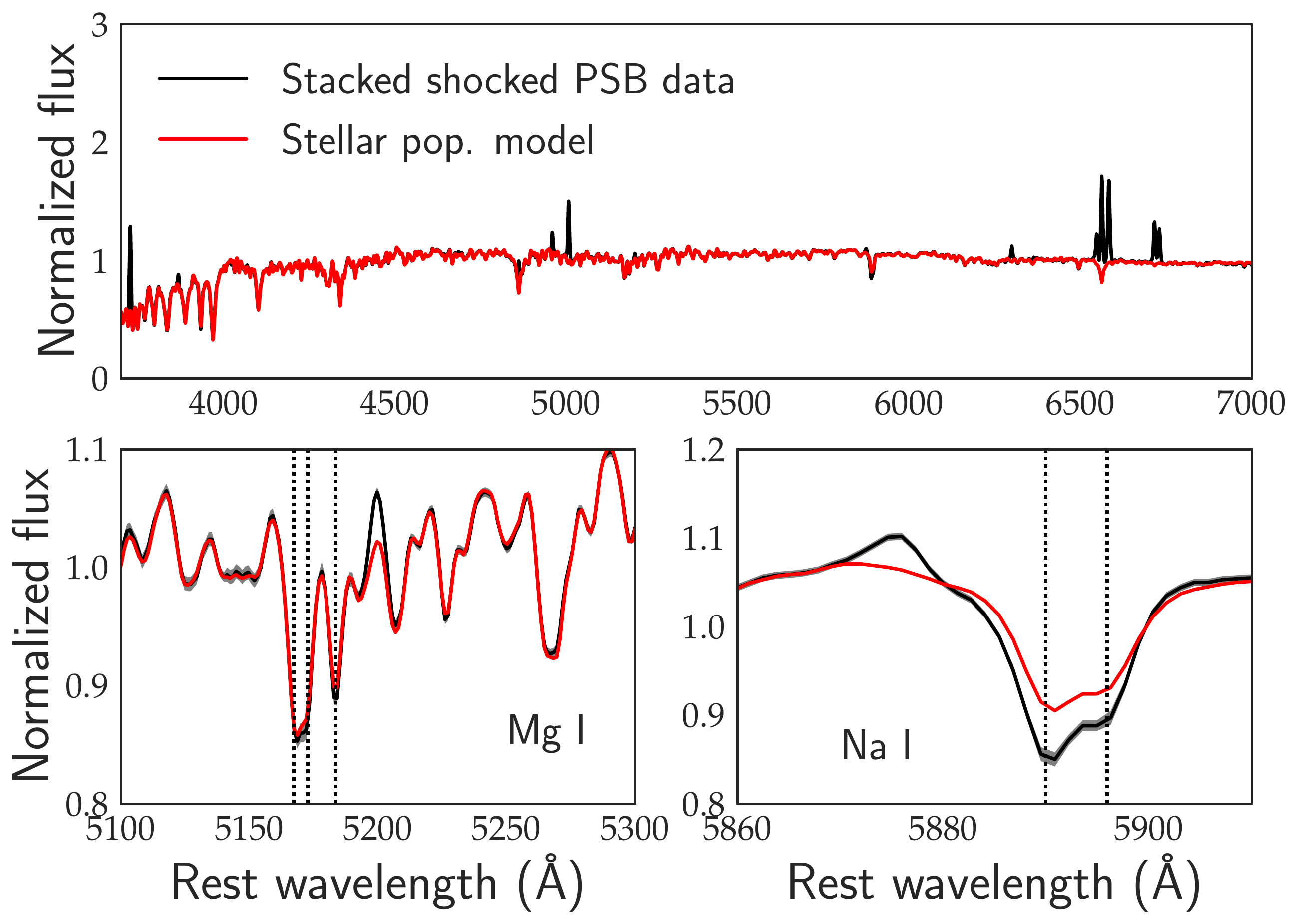}
\includegraphics[width=0.8\linewidth]{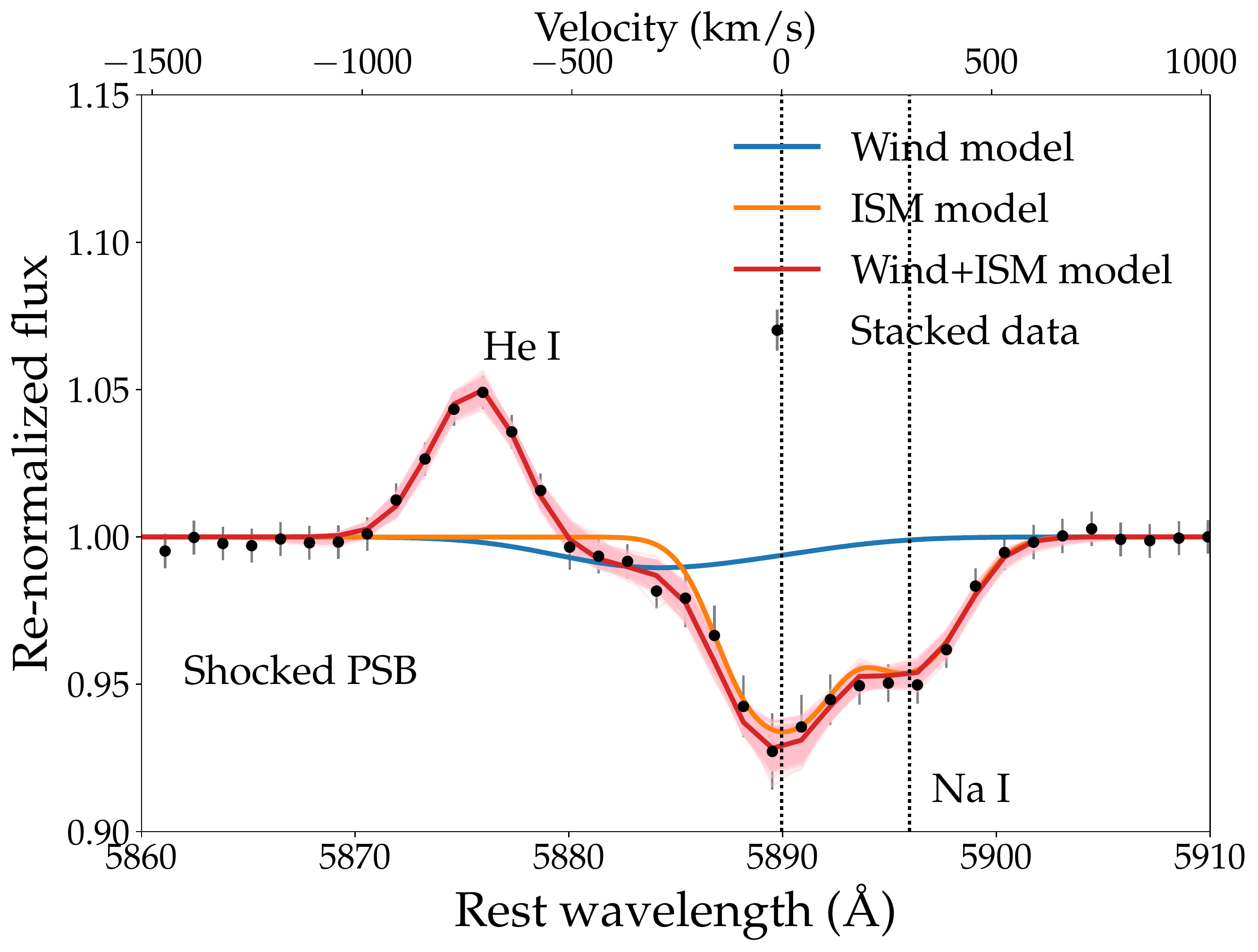} 
\caption{Similar to Figure~\ref{fig:agn_cont_spec}, but here the stacked spectrum of shocked post-starbursts studied by \citet{Alatalo+16a} are analyzed with a wind model after re-normalizing it with a stellar population synthesis model. Unlike the AGN PSB (Seyferts) in Figure~\ref{fig:agn_cont_spec}, the shocked PSB include LINERs and composite galaxies in the BPT AGN diagnostic.  \citet{Alatalo+16a} reported that the shocked PSBs have excess \ion{Na}{i}; here we attempt to provide insight on the source of  \ion{Na}{i} absorption in these galaxies. The estimated centroid and maximum velocities are $-355^{+140}_{-278} $and $-722^{+146}_{-298}$\,\kms respectively. The equivalent width of the ISM component is $0.67^{+0.04}_{-0.06}$\,{\AA} while that of the wind component is only $0.14^{+0.06}_{-0.04}$\,{\AA}. The wind velocity, and the wind and ISM equivalent widths are similar to those of AGN PSBs in Figure~\ref{fig:apsb_fit}. Details of the wind and ISM properties of the control sample are given in the text. \label{fig:spsb_fit}}
\end{figure*}

\begin{figure*}
\centering
\includegraphics[width=0.8\linewidth]{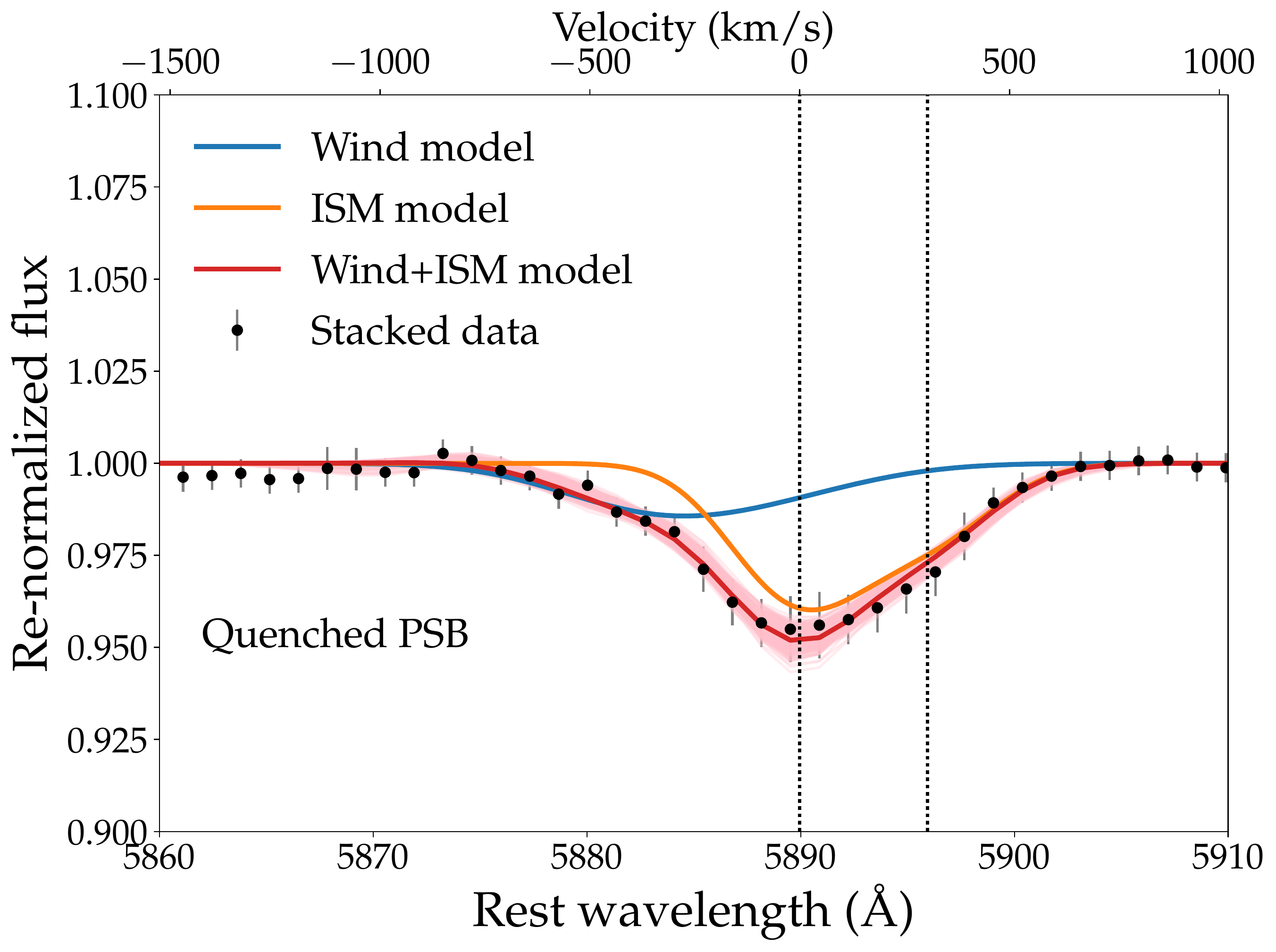}
\caption{The \ion{Na}{i} profile fit of the quenched PSB stacked spectrum shown in Figure~\ref{fig:qpsb_spec}. A persistent wind with a velocity of $-355^{+113}_{-98}$\,\kms is detected. As discussed in the text, the strong ISM component implies that $\sim 10^8-10^9$\,M$_\odot$ \ion{H}{i} gas may still be present in these recently quenched galaxies. \label{fig:qpsb_fit}}
\end{figure*}


\subsection{Simple Wind Model}

To fit the residual \ion{Na}{i} 5889.95, 5895.92\,{\AA} absorption profile after renormalizing by the stellar continuum fit, we adopt the partial covering wind model of \citet{Rupke+05a}. Similar models are used in recent works \citep{Sato+09, Chen+10,Rubin+14,Yesuf+17a}. Due to the limitations of observational data to fully constrain the wind model, the following simplifying assumptions were customarily made in previous works and are also used in this work: 1) The covering factor of the wind is independent of velocity. 2) The absorption-line profile shape is due entirely to the absorption of the stellar continuum. Scattered emission infill may, however, also affect the absorption profile. This effect is expected to be small because most galaxies in our samples have face-on orientation \citep{Chen+10}. 3) Two absorption components, an ISM component centered at zero velocity and a wind component, are sufficient to characterize the observed absorption-line profiles. This assumption may result in inaccurate column densities and line widths, if the profiles are composed of multi-components from multiple clouds but should usefully compare the strength and flow velocity of the \ion{Na}{i} absorption profile across different galaxies. 4) The velocity distribution of absorbing atoms within a component is Maxwellian, so that each absorption optical depth is modeled as a Gaussian $\tau(\lambda) = \tau_c {\rm e}^{-(\lambda-\lambda_c)^2/( \lambda_c\,b/c)^2}$ where  $\tau_c$  is a central optical depth at the line center ($\lambda_c$), $c$ is the speed of light, and $b=\sqrt2\,\sigma$ is the Doppler parameter. This assumption is likely to be an over-simplification, but is reasonable given that the observed shape of the absorption trough is strongly influenced by the instrumental resolution. 6) Following \citet{Chen+10},  the width of the ISM component (i.e., its Doppler parameter) is assumed to be the same as that of the of \ion{He}{i} 5876\,{\AA} emission-line. This line is, thus, fitted simultaneously with \ion{Na}{i} absorption-line, assuming both have a Gaussian shape. The covering fraction is known to be degenerate with the column density. We do not aim to constrain these two values independently.

With this simple model, the renormalized flux can be described as a product of the line intensity of the galaxy component and of the wind component. Each component has the form
 $1-C (1-{\rm e}^{-\tau(\lambda)})$. C is the covering fraction. One can express the centroid wavelength, $\lambda_c$, in terms of centroid velocity $v=c(\lambda_c-\lambda_0)/\lambda_0$, where $\lambda_0$ is the rest wavelength of the transition. For the galaxy component, $ \lambda_c=\lambda_0$ (i.e., no velocity shift). The central optical depth is expressed in terms of the column density ($N$), oscillator strength ($f_0$), and $b$ using the relation $\tau_c = 1.45 \times 10^{-15} \lambda_0\,f_0N/b$ for $N$ in units of cm$^{-2}$, $\lambda$ in {\AA} and $b$ in km\,s$^{-1}$. For \ion{Na}{i} $\lambda_0=5895.92$\,{\AA}, $f_0$ is 0.64 and for \ion{Na}{i} $\lambda_0=5889.95$\,{\AA} $f_0$ is 0.32.

The eight free parameters of this two component model are: the covering fractions of the wind ($C_w$) and of the ISM in the galaxy ( ${C_g}$), the velocity centroid shift of the wind ($v_{w}$), the Doppler parameters of the wind ($b_{w}$) and of the ISM ($b_g$), column densities of the wind ($N_{w}$), and of the gas in the galaxy ($N_{g}$), and the amplitude of the \ion{He}{i} 5876\,{\AA} emission-line ($A_g$). The doppler width of the  \ion{He}{i} line is assumed to be the same as that of the ISM. The model is convolved with the instrumental resolution and interpolated to the observed wavelength bin to match the observed data before comparing the two. 

The model is fit to the data using a Bayesian method with custom Python code \citep{Yesuf+17a}. The posterior probability densities (PDFs) of the model parameters were computed using the affine-invariant ensemble Metropolis-Hastings sampling algorithm \citep{Foreman-Mackey+13} assuming uniform priors: $C_w = C_g = (0.01,1)$, $v_{w} =(-1000, 250)$\,\kms, $b_{w} = b_{g} = (20, 300)$\,\kms, and $\log N_{w} = \log N_{g}=(12, 15.5)$ \cmtwo. Even though, for simplicity, we referred to $v_{w}$ as a wind velocity, a priori it can also be infall velocity. The priors encompass the ranges of previously measured wind parameters in local galaxies \citep[e.g.,][]{Rupke+05c,Martin+06,Chen+10}. To compute the likelihood of the data given the model parameters, we assumed that each data point is drawn from independent Gaussians centered around the model profile with a dispersion given by the measurement errors. This is equivalent to assuming a $\chi^2$ distribution for the sum of squares of normalized flux differences between the model and the data.

\subsection{Wind properties of AGN PSBs, Shocked PSBs, and Quenched PSBs}
The top panel of Figure~\ref{fig:agn_cont_spec} compares the stacked spectra of AGN PSBs and their control sample before they were re-normalized with a stellar population model fit to account for stellar absorption. The bottom panels of Figure~\ref{fig:agn_cont_spec} show the zoomed-in portion around \ion{Na}{i} before and after the re-normalization. The observed \ion{Na}{i} profiles of the two samples are similar but there is a significant difference at the high velocity tails. Figure~\ref{fig:apsb_fit} shows that, overall, the stacked spectrum of AGN PSBs are well fit by the stellar population model. The stellar population model, however, does not fully account for the observed \ion{Na}{i} absorption, which in addition is affected by the inter-stellar and wind absorptions. The bottom panel of Figure~\ref{fig:apsb_fit} shows the the residual \ion{Na}{i} profile, after re-normalizing by stellar population model fit, and the resulting fit of the ISM+Wind model. 

Figure~\ref{fig:apsb_cont_fit} is similar to Figure~\ref{fig:apsb_fit} but shows the corresponding plots for the control sample of AGN PSBs. Figure~\ref{fig:spsb_fit} shows similar plots for the shocked PSBs studied by \citet{Alatalo+16a}. Figure~\ref{fig:qpsb_fit} portrays the results of the ISM+Wind model for the quenched PSBs.

\begin{table*}
\caption{Wind model parameter fits to \ion{Na}{i} profiles. The parameters of the model are: the covering fractions of the wind ($C_w$) and of the ISM in the galaxy (${C_g}$), the velocity centroid shift of the wind ($v_{w}$), the Doppler parameters of the wind ($b_{w}=\sigma_w/\sqrt 2$) and of the ISM ($b_g$), column densities of the wind ($N_{w}$), and of the gas in the galaxy ($N_{g}$), and the amplitude of the \ion{He}{i} 5876\,{\AA} emission-line ($A_g$). The maximum velocity is defined as $v_w-2\sigma_w$. We use the notation $X^{+Y}_{-Z}$ to denote the median as X, the 84th percentile  as $X+Y$ and the 16th percentile as $X-Z$. \label{tab:fit}}
\begin{tabular}{lcccccccc} 
		\hline
		Model Parameters &AGN PSBs& $\longleftrightarrow $ & Control  & Shocked PSBs & $ \longleftrightarrow$& Control & Quenched PSBs \\
		\hline
		$C_w$ & $0.2^{+0.2}_{-0.1}$ &   & $0.09^{+0.17}_{-0.06}$ &$0.2^{+0.2}_{-0.1} $& &$0.1^{+0.1}_{-0.1} $&$0.2^{+0.2}_{-0.2} $ \\
		$v_w$ (\kms) & $-252^{+64}_{-57}$ & &$-119^{+33}_{-41}$&$-355^{+140}_{-278}$& &$-90^{+52}_{-119} $& $-355^{+113}_{-98} $ \\
		$b_w$ (\kms) & $307^{+28}_{-44}$ & &$206^{+31}_{-34}$&$288^{+45}_{-89}$& &$212^{+28}_{-67} $ & $324^{+20}_{-45} $ \\
		$\log N_w$ (\cmtwo) & $12.5^{+0.4}_{-0.2}$ & & $13.0^{+0.6}_{-0.5}$& $12.3^{+0.5}_{-0.2} $ & &$12.4^{+0.6}_{-0.3} $&$12.5^{+0.6}_{-0.3} $\\
		$C_g$ & $0.3^{+0.2}_{-0.1}$& & $0.2^{+0.2}_{-0.1}$ &$0.08^{+0.02}_{-0.01} $ & &$0.05^{+0.07}_{-0.01}$ & $0.2^{+0.2}_{-0.1} $\\
		$b_g$ (\kms) & $204^{+6}_{-7}$ & & $166^{+4}_{-4}$ & $152^{+11}_{-8} $ & & $145^{+2}_{-2} $ & $219^{+19}_{-21} $\\
		$\log N_g$ (\cmtwo) & $13.0^{+0.3}_{-0.2}$ & & $12.7^{+0.4}_{-0.3}$ &$13.6^{+0.1}_{-0.1} $& & $13.4^{+0.4}_{-0.3} $&$12.9^{+0.4}_{-0.3} $ \\
		$A_g$ & $0.076^{+0.002}_{-0.002}$ & &$0.078^{+0.002}_{-0.002}$ &$0.050^{+0.006}_{-0.003} $& &$0.115^{+0.001}_{-0.001} $& $0.002^{+0.003}_{-0.001} $ \\
		\hline
		Derived quantities & AGN PSBs& $\longleftrightarrow $ & Control  & Shocked PSBs & $ \longleftrightarrow$ &  Control & Quenched PSBs\\
		\hline
		$v_{\rm{max},w}$ (\kms) & $-678^{+54}_{-53}$  & &  $-406^{+51}_{-61}$ &$-722^{+146}_{-298} $& &$-402^{+49}_{-48}$ & $-778^{+88}_{-122} $ \\
		$\rm{Wind \,Eq.\, Width}$ (\AA) &$0.36^{+0.10}_{-0.07}$  & & $0.24^{+0.07}_{-0.06}$ &$0.14^{+0.06}_{-0.04} $& &$0.11^{+0..09}_{-0.06} $& $0.19^{+0.05}_{-0.04} $ \\
		$\rm{ISM \,Eq.\,Width}$ (\AA) & $0.65^{+0.07}_{-0.09}$  & & $0.30^{+0.06}_{-0.08}$ &$0.67^{+0.04}_{-0.06} $& &$0.24^{+0.06}_{-0.06} $& $0.44^{+0.04}_{-0.07} $ \\
		\hline
	\end{tabular}
\end{table*}

Table~\ref{tab:fit} summarizes the results of fitting the two component model to the residual \ion{Na}{i} profiles, after re-normalizing by stellar population model fit. The winds in AGN PSBs and shocked PSBs have significantly higher centroid velocities ($\gtrsim -250$\,\kms) and maximum velocities ($v_w-2\,\sigma_w \gtrsim -650$\,\kms,  where $\sigma_w= b_w/\sqrt 2$ is the velocity dispersion of the wind) than those in their control samples (centroid velocities $\sim -150$\,\kms and maximum velocities ($\sim -400$\,\kms). However, as we will discuss in detail in the next section, all winds are too weak to be associated with rapid quenching of star-formation in PSBs. The equivalent widths due to the winds are similar in AGN PSBs ($0.36^{+0.10}_{-0.07}$\,{\AA}) and in the control sample ($0.24^{+0.07}_{-0.06}$\,{\AA}) while the ISM contribution to the total equivalent widths is higher in the AGN ($0.65^{+0.07}_{-0.09}$\,{\AA}) than in the control sample ($0.30^{+0.06}_{-0.08}$ {\AA}). Similarly, for the shocked PSB sample of \citet{Alatalo+16a}, the ISM equivalent width is much higher than the wind equivalent width. The quenched PSBs, interestingly, have persistent winds similar in velocity to those in AGN PSBs. The wind equivalent width is $0.19^{+0.05}_{-0.04}$ {\AA}. It is also about half of the ISM component. 

The H$\alpha$/H$\beta$ ratios are not well measured in quenched PSBs, and their WISE 12\,$\mu$m to 4.6\,$\mu$m flux ratios are also not easy to interpret. So we tried to construct a control sample for quenched PSBs following section~\ref{sec:cont} but without these two quantities and we do not detect winds in the control sample.



\section{Discussion}\label{sec:disc}

We analyze the \ion{Na}{i} profile in stacked spectrum of 560 post-starburst (PSB) galaxies which exhibit emission-line ratios of AGN and in the stacked spectrum of their control sample. We  detect a wind with a centroid velocity of $-252^{+64}_{-57}$\,\kms and a maximum velocity of $-678^{+54}_{53}$\,\kms in the AGN PSBs. We detect a lower velocity wind in the control sample. We also detect winds in shocked PSBs \citep{Alatalo+16a} and quenched PSBs \citep[e.g.,][]{Goto07,Yesuf+14}. In this section, we discuss the \ion{H}{i} mass implied by the ISM component of \ion{Na}{i} profile and the wind mass outflow rate inferred from the wind component. As a comparison to the neutral gas outflow detected in \ion{Na}{i} absorption, we also discuss the ionized wind velocity and outflow rate detected using \ion{O}{iii} 4959, 5007\,{\AA} emission in the stacked spectrum AGN PSBs.

\subsection{The Inferred HI Gas Mass in PSBs is  $\sim 10^9\, M_\odot$}

There is no definitive observational evidence to constrain whether the quenching of star-formation in PSBs is driven by mechanisms such as AGN feedback, which results in complete removal or destruction of cold gas \citep{Hopkins+06} or other mechanisms such as morphological quenching \citep{Martig+09} that do not require such gas depletion. Existing observations indicate that some PSBs have significant amounts of cold gas \citep[e.g.,][]{Buyle+06,Zwaan+13,Rowlands+15,French+15,Alatalo+16b,Yesuf+17b}

In this section, we discuss the \ion{H}{i} mass estimate of post-starbursts based on the ISM component of \ion{Na}{i} absorption. We find a crude gas mass estimate of $\sim 0.6 - 6 \times 10^9\, M_\odot$ in PSBs, which is consistent with measurements of \citet{Zwaan+13}. The authors studied \ion{H}{i} gas in 11 quenched PSBs $z \sim 0.03$ and found that at least half of the PSBs have detectable \ion{H}{i} gas reservoirs. The \ion{H}{i}-detected PSBs have gas fractions similar to those observed in the most gas-rich early-type galaxies and the most gas-poor late-type galaxies \citep{Catinella+12}. For their sample, the mean \ion{H}{i} gas to stellar mass $M_\ion{H}{I}/M_\star \sim  0.03$ and $M_\ion{H}{i} \sim 2 \times 10^9$ M$_\odot$.

As shown in the equations below, the total mass of atomic hydrogen, $M_{\ion{H}{i}}$, in a PSB galaxy with radius $R_{g}$ can be estimated from the inferred \ion{Na}{i} column density, $N_{\ion{Na}{i}}$ , using the absorption-line modeling with some simplifying assumptions. Estimating the hydrogen gas column, $N_{\ion{H}{i}}$, from $N_{\ion{Na}{i}}$ requires knowing the abundance ratio of Na relative to hydrogen, $N_{\rm{Na}}/N_{\rm{H}}$,  the fraction of Na in gas phase, $f_{\rm{Na,gas}}$,  (i.e., not depleted onto dust grains), and the fraction of neutral sodium, $N_\ion{Na}{i}/N_{\rm{Na}}$.

\begin{equation}
\begin{aligned}
M_{\ion{H}{i}} & = m_p N_{\ion{H}{i}} \pi R_{g}^2 \\
& = 6.2 \times 10^8 M_\odot \left( \frac{N_\ion{H}{i}}{2.5 \times 10^{20} \, \rm{cm}^{-2}} \right) \left(\frac{R_{g}}{10\, \rm{kpc}}\right)^2
\end{aligned}
\end{equation}

\begin{equation}
N_{\ion{H}{i}}  = 2.5 \times 10^{20} \rm{cm}^{-2} \left(\frac{N_{\ion{Na}{i}}}{10^{13}}\right) \left(\frac{2 \times 10^{-6}}{N_{\rm{Na}}/N_{\rm{H}}}\right) \left(\frac{0.2}{N_\ion{Na}{i}/N_{\rm{Na}}}\right) \left(\frac{0.1}{f_{\rm{Na,gas}}}\right)
\end{equation}

\noindent where $m_p$ is the mass of a proton. The mean r-band Petrosian radius enclosing 90\% of the light is 9.8 kpc for our AGN PSBs. We assume a solar abundance ratio $N_{\rm{Na}}/N_{\rm{H}} = 2 \times 10^{-6}$ \citep{Savage+96}. Using the local mass-metallicity relation \citep{Tremonti+04}, for the typical galaxy mass in our sample, the mean oxygen gas abundance, 12 + log(O/H), of our sample is $\sim 9.1$ and the typical 1$\sigma$ scatter of the relation is $\sim 0.15$ dex. Therefore, the average gas phase metallicity of our sample is slightly higher than the assumed solar value of 8.7. This implies that the actual $N_{\rm{H}}$ is lower. In the cool ($10^3$ K) diffuse interstellar cloud toward $\zeta$ Oph in the Milky Way, the fraction of Na in gas phase is, $f_{\rm{Na,gas}} \sim 0.1$ \citep{Savage+96}. We assume a similar value for our galaxies. The first ionization potential of sodium is 5.1 eV and neutral \ion{Na}{i} is likely not the dominant ionization state. Even in infrared luminous Seyfert galaxies such as Mrk 273 and Arp 220, in which dust shielding is likely important, the fraction of neutral sodium is, $N_\ion{Na}{i}/N_{\rm{Na}} \sim 0.2$ \citep{Murray+07}. As the dusty starburst evolves, the $N_\ion{Na}{i}/N_{\rm{Na}}$ may become much smaller. If the $N_\ion{Na}{i}/N_{\rm{Na}}$ is between 0.02 -- 0.2, in PSBs, our estimated column densities in Table~\ref{tab:fit}, for the above assumptions, imply \ion{H}{i} mass of $\sim 0.6 - 6 \times 10^9\, M_\odot$ in AGN PSBs and quenched PSBs. In comparison, a typical, normal, quiescent galaxy with stellar mass of $5 \times 10^{10}\,M_\odot$ has an atomic gas of $\lesssim 2 \times 10^9\,M_\odot$ \citep{Catinella+12}. So, based on our estimate, it is very uncertain to conclude whether or not the PSBs have already achieved gas contents similar to quiescent galaxies. In general, the molecular hydrogen mass in a galaxy is about three times lower than the atomic hydrogen gas mass. So, our \ion{H}{i} mass estimate is consistent with the measured molecular gas mass range $\sim10^8-10^{10}$\,M$_\odot$ of PSBs \citep{Yesuf+17b,Alatalo+16b,French+15}. 

\subsection{Small Mass Outflow Rate in PSBs}
For a partially-filled wind of a thin shell radius $R_w$ subtending a solid angle $\Omega$, the mass of the wind can be expressed as \citep{Rupke+05b}: 
\begin{equation}
\begin{aligned}
M_{w} & = \Omega \mu m_p  N_{\ion{H}{i}} R_{w}^2 \\
& = 1.4 \times 10^8\, M_\odot \left(\frac{C_\Omega}{0.4}C_f \right)\left( \frac{N_\ion{H}{i}}{ 2.5 \times 10^{20} \, \rm{cm}^{-2}} \right) \left(\frac{R_{w}}{10\, \rm{kpc}}\right)^2
\end{aligned}
\end{equation}
where $\mu m_p$ is the mass per particle and $\mu=1.4$ is a correction for He abundance in the outflow and $\frac{\Omega}{4\pi} = C_\Omega C_f$.  $C_\Omega$ is the large-scale covering factor related to the opening angle of the wind, while $C_f$ is the local covering factor related to the clumpiness of the wind.

The mass outflow rate of such a wind is:

\begin{equation*}
\begin{aligned}
dM_w/dt  &  = M_w v_w/R_w \\
& = 3 \,M_\odot {\rm{yr}^{-1}} \left(\frac{C_\Omega}{0.4}C_f \right)\left( \frac{N_\ion{H}{i}}{ 2.5 \times 10^{20} \, \rm{cm}^{-2}} \right) \\
& \times \left(\frac{R_{w}}{10\, \rm{kpc}}\right)\left( \frac{v_w}{200\,{\rm km\, s^{-1}}}  \right)
\end{aligned}
\end{equation*}
For the wind parameter estimates in Table~\ref{tab:fit}, the mass outflow rate in AGN PSBs and quenched PSBs is $\sim 0.2\, \rm{M_\odot yr}^{-1}$. At this rate, within 1 Gyr, only a small fraction of the total \ion{H}{i} mass of a starburst would be expelled. In other words, the winds are too weak to have an effect on the gas in the starburst. Either the starburst has consumed most of the gas, or, perhaps, a quasar wind which we are not probing has already cleared out the gas, or substantial gas will remain in these galaxies. All we can firmly constrain at the moment is that  the low ionization neutral gas winds in these particular (AGN) PSBs are weak.

The mass outflow rate in the comparison sample is $\sim 0.1\rm{M_\odot yr}^{-1}$ and is smaller by about a factor of two. Given the substantial uncertainties in estimating the mass outflow rates, this difference is likely not significant. In the above estimate, we used the average wind velocity. But, note that even though the maximum velocity of AGN PSBs is higher than that of the control sample by about a factor of 1.7, using the maximum velocity instead does not increase the mass outflow estimate since the column density decreases exponentially and is lower by a factor of 7 at the maximum velocity.

Finally, we note that our assumption of a 10 kpc scale wind is not unreasonable since \citet{Martin+06} has found \ion{Na}{i} absorption that extends 4 --18 kpc in several ultra-luminous infrared galaxies  (ULIRGs). \citet{Rupke+15} have reported a ULIRG that has \ion{Na}{i} wind emission that extends at least 3 kpc.
 
\subsection{Similarly Low, Ionized-Gas Outflow Rate is Detected in Emission in AGN PSBs}

\begin{figure*}
\centering
\includegraphics[width=0.85\linewidth]{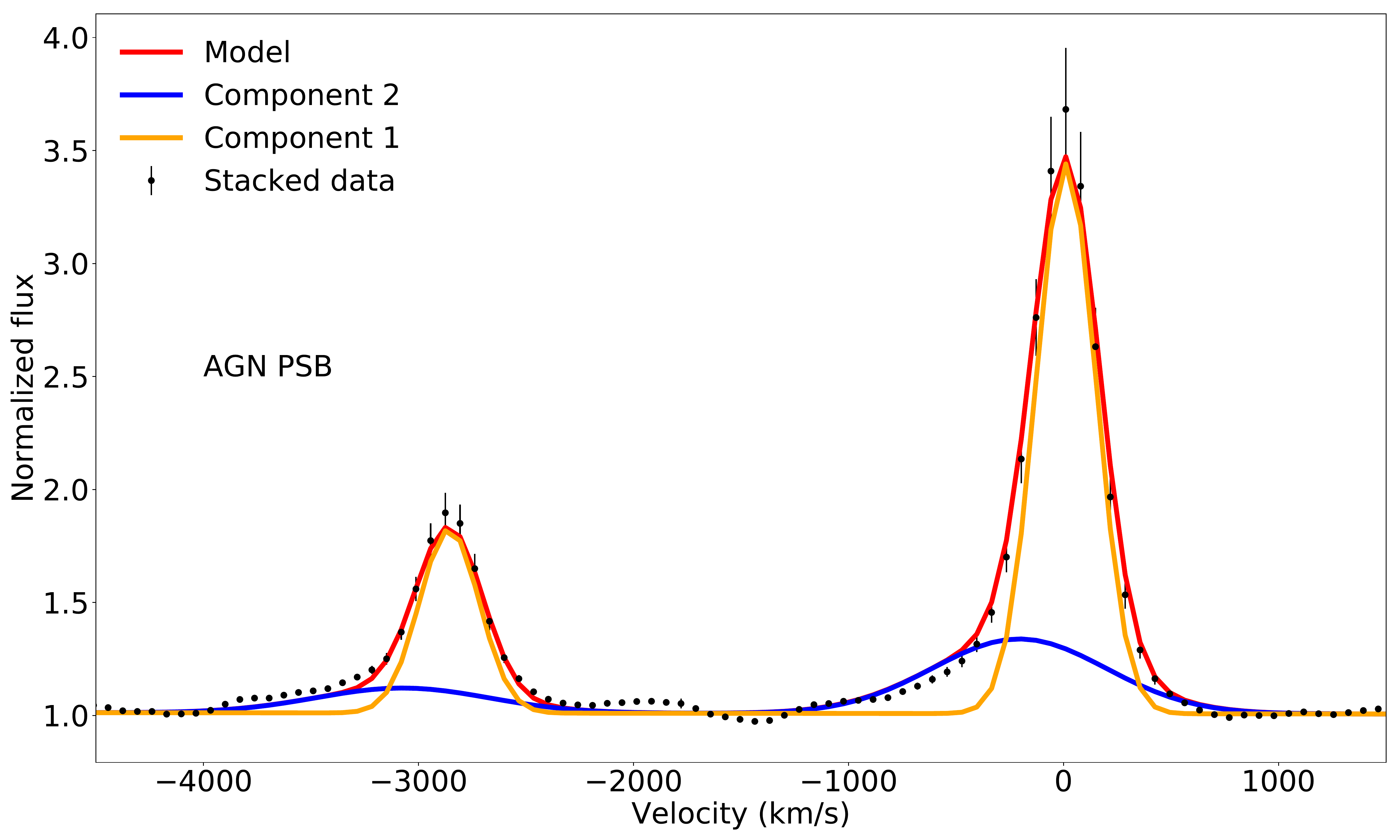}
\caption{The \ion{O}{iii} 4959, 5007 doublet is simultaneously fit with two component Gaussians. The doublet width for each component is the same and its amplitude is fixed to 2.99:1 ratio. The component shown in orange is consistent with zero shift while the one shown in blue is shifted by $\sim -200$\,\kms and has a velocity dispersion of $\sim 400$\,\kms. \label{fig:o3_fit}}
\end{figure*}

As discussed in detail in \citet{Yesuf+17b}, AGN winds detected in emission lines of ionized gas have been extensively studied \citep[e.g.,][]{ Liu+13,Harrison+14,Zakamska+14,McElroy+15,Harrison+16}. But, only very few works have studied AGN winds in both emission and absorption in the  same galaxies \citep{Rupke+13}. PSBs have not been studied this way at all.  \cite{Baron+17} recently reported the first evidence of AGN-driven wind using nebular emission-lines in a quiescent post-starburst galaxy.  But, this galaxy does not have \ion{Na}{i} absorption. The authors measured blue-shifted emission-lines with a mean $-130$\,\kms, with a velocity dispersion of $\sim 570$\,\kms, and a mass outflow rate of at least 4 $\rm{M_\odot yr}^{-1}$. The wind properties of this PSB are very different from our absorption-line based inference.  Specifically, our modeling results in a velocity dispersion of $\sim 190$\,\kms and an inferred outflow rate of $\sim 0.2\rm{M_\odot yr}^{-1}$. One may wonder whether the Baron object is very unusual -- after all, this object was found as an outlier galaxy by an anomaly detection algorithm -- or if the wind properties in absorption and emission are different. 
Similar to \citet{Harrison+14}, we simultaneously fit the \ion{O}{iii} doublets with two Gaussian components (for each doublet member) + a linear continuum (slope and intercept). The centroid of the Gaussian for each component is constrianed to shift by the same amount for the doublet. The width of the doublet for a given component is kept the same, and the flux ratio of the doublet was fixed to be 2.99. The model has eight free parameters: 3 parameters for each Gaussian component characterizing its amplitude, centroid shift and dispersion, and the slope and the intercept of the line. The model is fit using Levenberg-Marquardt least square minimization (using the Python \textbf{lmfit} package) to determine the parameters that minimize the difference between the model and data normalized by the bootstrap errors.

One of the Gaussian components has a velocity shift consistent with zero while the other has a velocity shift of $-205 \pm 54 $\,\kms and the Doppler parameter of $572 \pm 53$\,\kms (or a velocity dispersion of $404 \pm 37$\,\kms). While the emission-line velocity shift is consistent with absorption-line centroid velocity shift ($-252^{+64}_{-57}$\,\kms), its Doppler parameter is significantly higher than that of the \ion{Na}{i} absorption-line ($307^{+28}_{-44}\,$\,\kms). The Doppler parameter for the static component is $197 \pm 23 $\,\kms, which, as expected, is comparable to the velocity dispersion of the ISM component ($204^{+6}_{-7}$\,\kms) of the absorption-line model. The halo escape velocity is about six times the velocity dispersion \citep{Weiner+09}. Therefore, most of the wind mass detected either in emission or absorption cannot escape. The relative amplitude of \ion{O}{iii}\,5007 {\AA} for the static component is $2.4 \pm 0.4$ and for the wind component is $0.3 \pm 0.1$. The quality of the data does not allow a similar two component analysis to be conducted for the control sample or shocked PSBs and their control sample.

Assuming solar metallicity, and that most of the oxygen in the ionized outflow is in the form of O$^{+2}$, the mass outflow rate of the ionized spherical outflow can be estimated from \ion{O}{iii}\,5007 {\AA} emission-line luminosity \citep{Cano-Diaz+12}:
\begin{equation}
M_w = 5.7 \times 10^7\, M_\odot \left(\frac{C}{1} \right)\left(\frac{L_{\ion{O}{iii}}}{10^{44} \rm{erg\,s^{-1}}}\right)\left(\frac{10^3 \rm{cm^{-3}}}{\langle n_e\rangle}\right)
\end{equation}

\begin{equation}
\begin{aligned}
dM_w/dt  &  = 3M_w v_w/R_w \\
                & = 164\,M_\odot \, {\rm {yr}^{-1}} \left(\frac{C}{1} \right)\left(\frac{L_{\ion{O}{iii}}}{10^{44} \,\rm{erg\,s^{-1}}}\right)\left(\frac{10^3\, \rm{cm^{-3}}}{\langle n_e\rangle}\right) \\
                & \times \left(\frac{v_w}{10^3\,\rm{km\,s^{-1}}}\right)\left(\frac{1\,\rm{kpc}}{R_w}\right)
\end{aligned}
\end{equation}
where $\langle n_e\rangle$ is the volume averaged electron density and it is on the order of $10^3$ cm$^{-3}$ in AGN \citep[e.g.,][]{Perna+17}, and $C = \langle n_e\rangle^2/\langle n_e^2\rangle$. The mean total \ion{O}{iii} 5007 luminosity of our sample is 
$L_{\ion{O}{iii}} = 6 \times 10^{40}$ erg s$^{-1}$, and less than half ($\frac{0.3\times 572}{2.4\times 197}= 0.36$) of this luminosity is due to the wind. Even if we take the average velocity to be $v_w = -205-404 =-609$\,\kms, the wind radius to be 1 kpc, and $\langle n_e \rangle$ to be 100 cm$^{-3}$, the ionized mass outflow rate is still $\lesssim 0.3  \rm{M_\odot \, yr}^{-1} $. Therefore, it is very small and is not inconsistent with our \ion{Na}{i} absorption based estimate.

\section{Conclusions}\label{sec:conc}
Using a stellar population synthesis model and a simple ISM+wind model, we analyze the \ion{Na}{i} absorption profiles in the stacked SDSS spectra of several hundred post-starburst (PSB) galaxies at $z \sim 0.1$. Our main conclusions are the following:

\begin{itemize}

\item In the stacked spectrum of AGN PSBs, we detect a wind with a centroid velocity of $-252^{+64}_{-57}$\,\kms and a maximum velocity of $-678^{+54}_{-53}$\,\kms. The equivalent width of the wind is $0.36^{+0.10}_{-0.07}$ {\AA}, about a factor of two smaller than the equivalent width due to the ISM absorption in the host galaxy.
\smallskip
\item Statistically, the centroid wind velocities in AGN PSBs are about two times higher than those in the control sample, which are matched in stellar mass, axis-ratio, redshift, 4000 {\AA} break, H$\alpha$ to H$\beta$ and WISE 12\,$\mu$m to 4.6\,$\mu$m flux ratios. The control sample has a wind with a centroid velocity of $-119^{+33}_{-41}$\,\kms, a maximum velocity of $-406^{+51}_{-61}$\,\kms, and a wind strength of $0.24^{+0.07}_{-0.06}$ {\AA}.
\smallskip
\item We confirm the excess \ion{Na}{i} absorption in shocked PSBs reported in \citet{Alatalo+16a}. But, most of the absorption is likely due to ISM, not to wind absorption. The centroid wind velocity is $-355^{+140}_{-278}$\,\kms and the maximum velocity is $-722^{+146}_{-298}$\,\kms.
\smallskip
\item We detect a wind in the stacked spectrum of quenched PSBs, which has a centroid velocity of $-355^{+113}_{-98}$\,\kms, a maximum velocity of $-778^{+88}_{-122}$\,\kms, and a wind strength of $0.19^{+0.05}_{-0.04}$ {\AA}.\\

\end{itemize}

Despite the significant difference in the wind velocities of AGN PSBs and that of their control sample, the winds in these AGNs are estimated to be too weak to remove most of the cold gas in their host galaxies and thereby cause rapid quenching. Future high signal-to-noise individual spectra of AGN PSBs will be useful to confirm our results and further investigate the dependence of wind properties on AGN strength. \\

\bigskip

The authors gratefully acknowledge support from the National Science Foundation (NSF) under grant number AST-1615730. 

Funding for SDSS-III has been provided by the Alfred P. Sloan Foundation, the Participating Institutions, the National Science Foundation, and the U.S. Department of Energy Office of Science. The SDSS-III web site is http://www.sdss3.org/.

SDSS-III is managed by the Astrophysical Research Consortium for the Participating Institutions of the SDSS-III Collaboration including the University of Arizona, the Brazilian Participation Group, Brookhaven National Laboratory, Carnegie Mellon University, University of Florida, the French Participation Group, the German Participation Group, Harvard University, the Instituto de Astrofisica de Canarias, the Michigan State/Notre Dame/JINA Participation Group, Johns Hopkins University, Lawrence Berkeley National Laboratory, Max Planck Institute for Astrophysics, Max Planck Institute for Extraterrestrial Physics, New Mexico State University, New York University, Ohio State University, Pennsylvania State University, University of Portsmouth, Princeton University, the Spanish Participation Group, University of Tokyo, University of Utah, Vanderbilt University, University of Virginia, University of Washington, and Yale University.

\bibliographystyle{mnras}
\bibliography{reference.bib}

\begin{thebibliography}{}
\makeatletter
\relax
\def\mn@urlcharsother{\let\do\@makeother \do\$\do\&\do\#\do\^\do\_\do\%\do\~}
\def\mn@doi{\begingroup\mn@urlcharsother \@ifnextchar [ {\mn@doi@}
  {\mn@doi@[]}}
\def\mn@doi@[#1]#2{\def\@tempa{#1}\ifx\@tempa\@empty \href
  {http://dx.doi.org/#2} {doi:#2}\else \href {http://dx.doi.org/#2} {#1}\fi
  \endgroup}
\def\mn@eprint#1#2{\mn@eprint@#1:#2::\@nil}
\def\mn@eprint@arXiv#1{\href {http://arxiv.org/abs/#1} {{\tt arXiv:#1}}}
\def\mn@eprint@dblp#1{\href {http://dblp.uni-trier.de/rec/bibtex/#1.xml}
  {dblp:#1}}
\def\mn@eprint@#1:#2:#3:#4\@nil{\def\@tempa {#1}\def\@tempb {#2}\def\@tempc
  {#3}\ifx \@tempc \@empty \let \@tempc \@tempb \let \@tempb \@tempa \fi \ifx
  \@tempb \@empty \def\@tempb {arXiv}\fi \@ifundefined
  {mn@eprint@\@tempb}{\@tempb:\@tempc}{\expandafter \expandafter \csname
  mn@eprint@\@tempb\endcsname \expandafter{\@tempc}}}

\bibitem[\protect\citeauthoryear{{Alam} et~al.,}{{Alam} et~al.}{2015}]{Alam+15}
{Alam} S.,  et~al., 2015, \mn@doi [\apjs] {10.1088/0067-0049/219/1/12}, \href
  {http://adsabs.harvard.edu/abs/2015ApJS..219...12A} {219, 12}

\bibitem[\protect\citeauthoryear{{Alatalo} et~al.,}{{Alatalo}
  et~al.}{2016a}]{Alatalo+16a}
{Alatalo} K.,  et~al., 2016a, \mn@doi [\apjs] {10.3847/0067-0049/224/2/38},
  \href {http://adsabs.harvard.edu/abs/2016ApJS..224...38A} {224, 38}

\bibitem[\protect\citeauthoryear{{Alatalo} et~al.,}{{Alatalo}
  et~al.}{2016b}]{Alatalo+16b}
{Alatalo} K.,  et~al., 2016b, \mn@doi [\apj] {10.3847/0004-637X/827/2/106},
  \href {http://adsabs.harvard.edu/abs/2016ApJ...827..106A} {827, 106}

\bibitem[\protect\citeauthoryear{{Baldwin}, {Phillips}  \&
  {Terlevich}}{{Baldwin} et~al.}{1981}]{Baldwin+81}
{Baldwin} J.~A.,  {Phillips} M.~M.,   {Terlevich} R.,  1981, \mn@doi [\pasp]
  {10.1086/130766}, \href {http://adsabs.harvard.edu/abs/1981PASP...93....5B}
  {93, 5}

\bibitem[\protect\citeauthoryear{{Barnes} \& {Hernquist}}{{Barnes} \&
  {Hernquist}}{1991}]{Barnes+91}
{Barnes} J.~E.,  {Hernquist} L.~E.,  1991, \mn@doi [\apjl] {10.1086/185978},
  \href {http://adsabs.harvard.edu/abs/1991ApJ...370L..65B} {370, L65}

\bibitem[\protect\citeauthoryear{{Baron}, {Netzer}, {Poznanski}, {Prochaska}
  \& {Forster Schreiber}}{{Baron} et~al.}{2017}]{Baron+17}
{Baron} D.,  {Netzer} H.,  {Poznanski} D.,  {Prochaska} J.~X.,   {Forster
  Schreiber} N.~M.,  2017, preprint, \href
  {http://adsabs.harvard.edu/abs/2017arXiv170503891B} {} (\mn@eprint {arXiv}
  {1705.03891})

\bibitem[\protect\citeauthoryear{{Bekki}, {Couch}, {Shioya}  \&
  {Vazdekis}}{{Bekki} et~al.}{2005}]{Bekki+05}
{Bekki} K.,  {Couch} W.~J.,  {Shioya} Y.,   {Vazdekis} A.,  2005, \mn@doi
  [\mnras] {10.1111/j.1365-2966.2005.08932.x}, \href
  {http://adsabs.harvard.edu/abs/2005MNRAS.359..949B} {359, 949}

\bibitem[\protect\citeauthoryear{{Brinchmann}, {Charlot}, {White}, {Tremonti},
  {Kauffmann}, {Heckman}  \& {Brinkmann}}{{Brinchmann}
  et~al.}{2004}]{Brinchmann+04}
{Brinchmann} J.,  {Charlot} S.,  {White} S.~D.~M.,  {Tremonti} C.,  {Kauffmann}
  G.,  {Heckman} T.,   {Brinkmann} J.,  2004, \mn@doi [\mnras]
  {10.1111/j.1365-2966.2004.07881.x}, \href
  {http://adsabs.harvard.edu/abs/2004MNRAS.351.1151B} {351, 1151}

\bibitem[\protect\citeauthoryear{{Buyle}, {Michielsen}, {De Rijcke}, {Pisano},
  {Dejonghe}  \& {Freeman}}{{Buyle} et~al.}{2006}]{Buyle+06}
{Buyle} P.,  {Michielsen} D.,  {De Rijcke} S.,  {Pisano} D.~J.,  {Dejonghe} H.,
    {Freeman} K.,  2006, \mn@doi [\apj] {10.1086/505633}, \href
  {http://adsabs.harvard.edu/abs/2006ApJ...649..163B} {649, 163}

\bibitem[\protect\citeauthoryear{{Cano-D{\'{\i}}az}, {Maiolino}, {Marconi},
  {Netzer}, {Shemmer}  \& {Cresci}}{{Cano-D{\'{\i}}az}
  et~al.}{2012}]{Cano-Diaz+12}
{Cano-D{\'{\i}}az} M.,  {Maiolino} R.,  {Marconi} A.,  {Netzer} H.,  {Shemmer}
  O.,   {Cresci} G.,  2012, \mn@doi [\aap] {10.1051/0004-6361/201118358}, \href
  {http://adsabs.harvard.edu/abs/2012A%26A...537L...8C} {537, L8}

\bibitem[\protect\citeauthoryear{{Cappellari}}{{Cappellari}}{2017}]{Cappellari+17}
{Cappellari} M.,  2017, \mn@doi [\mnras] {10.1093/mnras/stw3020}, \href
  {http://adsabs.harvard.edu/abs/2017MNRAS.466..798C} {466, 798}

\bibitem[\protect\citeauthoryear{{Cappellari} \& {Emsellem}}{{Cappellari} \&
  {Emsellem}}{2004}]{Cappellari+04}
{Cappellari} M.,  {Emsellem} E.,  2004, \mn@doi [\pasp] {10.1086/381875}, \href
  {http://adsabs.harvard.edu/abs/2004PASP..116..138C} {116, 138}

\bibitem[\protect\citeauthoryear{{Catinella} et~al.,}{{Catinella}
  et~al.}{2012}]{Catinella+12}
{Catinella} B.,  et~al., 2012, \mn@doi [\aap] {10.1051/0004-6361/201219261},
  \href {http://adsabs.harvard.edu/abs/2012A%26A...544A..65C} {544, A65}

\bibitem[\protect\citeauthoryear{{Chen}, {Tremonti}, {Heckman}, {Kauffmann},
  {Weiner}, {Brinchmann}  \& {Wang}}{{Chen} et~al.}{2010}]{Chen+10}
{Chen} Y.-M.,  {Tremonti} C.~A.,  {Heckman} T.~M.,  {Kauffmann} G.,  {Weiner}
  B.~J.,  {Brinchmann} J.,   {Wang} J.,  2010, \mn@doi [\aj]
  {10.1088/0004-6256/140/2/445}, \href
  {http://adsabs.harvard.edu/abs/2010AJ....140..445C} {140, 445}

\bibitem[\protect\citeauthoryear{{Cimatti} et~al.,}{{Cimatti}
  et~al.}{2013}]{Cimatti+13}
{Cimatti} A.,  et~al., 2013, \mn@doi [\apjl] {10.1088/2041-8205/779/1/L13},
  \href {http://adsabs.harvard.edu/abs/2013ApJ...779L..13C} {779, L13}

\bibitem[\protect\citeauthoryear{{Coil}, {Weiner}, {Holz}, {Cooper}, {Yan}  \&
  {Aird}}{{Coil} et~al.}{2011}]{Coil+11}
{Coil} A.~L.,  {Weiner} B.~J.,  {Holz} D.~E.,  {Cooper} M.~C.,  {Yan} R.,
  {Aird} J.,  2011, \mn@doi [\apj] {10.1088/0004-637X/743/1/46}, \href
  {http://adsabs.harvard.edu/abs/2011ApJ...743...46C} {743, 46}

\bibitem[\protect\citeauthoryear{{Di Matteo}, {Springel}  \& {Hernquist}}{{Di
  Matteo} et~al.}{2005}]{DiMatteo+05}
{Di Matteo} T.,  {Springel} V.,   {Hernquist} L.,  2005, \mn@doi [\nat]
  {10.1038/nature03335}, \href
  {http://adsabs.harvard.edu/abs/2005Natur.433..604D} {433, 604}

\bibitem[\protect\citeauthoryear{{Diamond-Stanic}, {Moustakas}, {Tremonti},
  {Coil}, {Hickox}, {Robaina}, {Rudnick}  \& {Sell}}{{Diamond-Stanic}
  et~al.}{2012}]{Diamond-stanic+12}
{Diamond-Stanic} A.~M.,  {Moustakas} J.,  {Tremonti} C.~A.,  {Coil} A.~L.,
  {Hickox} R.~C.,  {Robaina} A.~R.,  {Rudnick} G.~H.,   {Sell} P.~H.,  2012,
  \mn@doi [\apjl] {10.1088/2041-8205/755/2/L26}, \href
  {http://adsabs.harvard.edu/abs/2012ApJ...755L..26D} {755, L26}

\bibitem[\protect\citeauthoryear{{Donoso} et~al.,}{{Donoso}
  et~al.}{2012}]{Donoso+12}
{Donoso} E.,  et~al., 2012, \mn@doi [\apj] {10.1088/0004-637X/748/2/80}, \href
  {http://adsabs.harvard.edu/abs/2012ApJ...748...80D} {748, 80}

\bibitem[\protect\citeauthoryear{{Dressler} \& {Gunn}}{{Dressler} \&
  {Gunn}}{1983}]{Dressler+83}
{Dressler} A.,  {Gunn} J.~E.,  1983, \mn@doi [\apj] {10.1086/161093}, \href
  {http://adsabs.harvard.edu/abs/1983ApJ...270....7D} {270, 7}

\bibitem[\protect\citeauthoryear{{Foreman-Mackey}, {Hogg}, {Lang}  \&
  {Goodman}}{{Foreman-Mackey} et~al.}{2013}]{Foreman-Mackey+13}
{Foreman-Mackey} D.,  {Hogg} D.~W.,  {Lang} D.,   {Goodman} J.,  2013, \mn@doi
  [\pasp] {10.1086/670067}, \href
  {http://adsabs.harvard.edu/abs/2013PASP..125..306F} {125, 306}

\bibitem[\protect\citeauthoryear{{French}, {Yang}, {Zabludoff}, {Narayanan},
  {Shirley}, {Walter}, {Smith}  \& {Tremonti}}{{French}
  et~al.}{2015}]{French+15}
{French} K.~D.,  {Yang} Y.,  {Zabludoff} A.,  {Narayanan} D.,  {Shirley} Y.,
  {Walter} F.,  {Smith} J.-D.,   {Tremonti} C.~A.,  2015, \mn@doi [\apj]
  {10.1088/0004-637X/801/1/1}, \href
  {http://adsabs.harvard.edu/abs/2015ApJ...801....1F} {801, 1}

\bibitem[\protect\citeauthoryear{{Geach} et~al.,}{{Geach}
  et~al.}{2014}]{Geach+14}
{Geach} J.~E.,  et~al., 2014, \mn@doi [\nat] {10.1038/nature14012}, \href
  {http://adsabs.harvard.edu/abs/2014Natur.516...68G} {516, 68}

\bibitem[\protect\citeauthoryear{{Goto}}{{Goto}}{2007}]{Goto07}
{Goto} T.,  2007, \mn@doi [\mnras] {10.1111/j.1365-2966.2007.12227.x}, \href
  {http://adsabs.harvard.edu/abs/2007MNRAS.381..187G} {381, 187}

\bibitem[\protect\citeauthoryear{{Harrison}, {Alexander}, {Mullaney}  \&
  {Swinbank}}{{Harrison} et~al.}{2014}]{Harrison+14}
{Harrison} C.~M.,  {Alexander} D.~M.,  {Mullaney} J.~R.,   {Swinbank} A.~M.,
  2014, \mn@doi [\mnras] {10.1093/mnras/stu515}, \href
  {http://adsabs.harvard.edu/abs/2014MNRAS.441.3306H} {441, 3306}

\bibitem[\protect\citeauthoryear{{Harrison} et~al.,}{{Harrison}
  et~al.}{2016}]{Harrison+16}
{Harrison} C.~M.,  et~al., 2016, \mn@doi [\mnras] {10.1093/mnras/stv2727},
  \href {http://adsabs.harvard.edu/abs/2016MNRAS.456.1195H} {456, 1195}

\bibitem[\protect\citeauthoryear{{Hayward}, {Torrey}, {Springel}, {Hernquist}
  \& {Vogelsberger}}{{Hayward} et~al.}{2014}]{Hayward+14}
{Hayward} C.~C.,  {Torrey} P.,  {Springel} V.,  {Hernquist} L.,
  {Vogelsberger} M.,  2014, \mn@doi [\mnras] {10.1093/mnras/stu957}, \href
  {http://adsabs.harvard.edu/abs/2014MNRAS.442.1992H} {442, 1992}

\bibitem[\protect\citeauthoryear{{Hopkins}, {Hernquist}, {Cox}, {Di Matteo},
  {Robertson}  \& {Springel}}{{Hopkins} et~al.}{2006}]{Hopkins+06}
{Hopkins} P.~F.,  {Hernquist} L.,  {Cox} T.~J.,  {Di Matteo} T.,  {Robertson}
  B.,   {Springel} V.,  2006, \mn@doi [\apjs] {10.1086/499298}, \href
  {http://adsabs.harvard.edu/abs/2006ApJS..163....1H} {163, 1}

\bibitem[\protect\citeauthoryear{{Hopkins}, {Hernquist}, {Cox}  \& {Kere{\v
  s}}}{{Hopkins} et~al.}{2008}]{Hopkins+08}
{Hopkins} P.~F.,  {Hernquist} L.,  {Cox} T.~J.,   {Kere{\v s}} D.,  2008,
  \mn@doi [\apjs] {10.1086/524362}, \href
  {http://adsabs.harvard.edu/abs/2008ApJS..175..356H} {175, 356}

\bibitem[\protect\citeauthoryear{{Ishibashi} \& {Fabian}}{{Ishibashi} \&
  {Fabian}}{2016}]{Ishibashi+16}
{Ishibashi} W.,  {Fabian} A.~C.,  2016, \mn@doi [\mnras]
  {10.1093/mnras/stw2063}, \href
  {http://adsabs.harvard.edu/abs/2016MNRAS.463.1291I} {463, 1291}

\bibitem[\protect\citeauthoryear{{Kauffmann} et~al.,}{{Kauffmann}
  et~al.}{2003}]{kauffmann03c}
{Kauffmann} G.,  et~al., 2003, \mn@doi [\mnras]
  {10.1111/j.1365-2966.2003.07154.x}, \href
  {http://adsabs.harvard.edu/abs/2003MNRAS.346.1055K} {346, 1055}

\bibitem[\protect\citeauthoryear{{Kewley}, {Dopita}, {Sutherland}, {Heisler}
  \& {Trevena}}{{Kewley} et~al.}{2001}]{Kewley+01}
{Kewley} L.~J.,  {Dopita} M.~A.,  {Sutherland} R.~S.,  {Heisler} C.~A.,
  {Trevena} J.,  2001, \mn@doi [\apj] {10.1086/321545}, \href
  {http://adsabs.harvard.edu/abs/2001ApJ...556..121K} {556, 121}

\bibitem[\protect\citeauthoryear{{Krug}, {Rupke}  \& {Veilleux}}{{Krug}
  et~al.}{2010}]{Krug+10}
{Krug} H.~B.,  {Rupke} D.~S.~N.,   {Veilleux} S.,  2010, \mn@doi [\apj]
  {10.1088/0004-637X/708/2/1145}, \href
  {http://adsabs.harvard.edu/abs/2010ApJ...708.1145K} {708, 1145}

\bibitem[\protect\citeauthoryear{{Liu}, {Zakamska}, {Greene}, {Nesvadba}  \&
  {Liu}}{{Liu} et~al.}{2013}]{Liu+13}
{Liu} G.,  {Zakamska} N.~L.,  {Greene} J.~E.,  {Nesvadba} N.~P.~H.,   {Liu} X.,
   2013, \mn@doi [\mnras] {10.1093/mnras/stt1755}, \href
  {http://adsabs.harvard.edu/abs/2013MNRAS.436.2576L} {436, 2576}

\bibitem[\protect\citeauthoryear{{Martig}, {Bournaud}, {Teyssier}  \&
  {Dekel}}{{Martig} et~al.}{2009}]{Martig+09}
{Martig} M.,  {Bournaud} F.,  {Teyssier} R.,   {Dekel} A.,  2009, \mn@doi
  [\apj] {10.1088/0004-637X/707/1/250}, \href
  {http://adsabs.harvard.edu/abs/2009ApJ...707..250M} {707, 250}

\bibitem[\protect\citeauthoryear{{Martin}}{{Martin}}{2006}]{Martin+06}
{Martin} C.~L.,  2006, \mn@doi [\apj] {10.1086/504886}, \href
  {http://adsabs.harvard.edu/abs/2006ApJ...647..222M} {647, 222}

\bibitem[\protect\citeauthoryear{{McElroy}, {Croom}, {Pracy}, {Sharp}, {Ho}  \&
  {Medling}}{{McElroy} et~al.}{2015}]{McElroy+15}
{McElroy} R.,  {Croom} S.~M.,  {Pracy} M.,  {Sharp} R.,  {Ho} I.-T.,
  {Medling} A.~M.,  2015, \mn@doi [\mnras] {10.1093/mnras/stu2224}, \href
  {http://adsabs.harvard.edu/abs/2015MNRAS.446.2186M} {446, 2186}

\bibitem[\protect\citeauthoryear{{Murray}, {Martin}, {Quataert}  \&
  {Thompson}}{{Murray} et~al.}{2007}]{Murray+07}
{Murray} N.,  {Martin} C.~L.,  {Quataert} E.,   {Thompson} T.~A.,  2007,
  \mn@doi [\apj] {10.1086/512660}, \href
  {http://adsabs.harvard.edu/abs/2007ApJ...660..211M} {660, 211}

\bibitem[\protect\citeauthoryear{{Nedelchev}, {Sarzi}  \&
  {Kaviraj}}{{Nedelchev} et~al.}{2017}]{Nedelchev+17}
{Nedelchev} B.,  {Sarzi} M.,   {Kaviraj} S.,  2017, preprint, \href
  {http://adsabs.harvard.edu/abs/2017arXiv170507994N} {} (\mn@eprint {arXiv}
  {1705.07994})

\bibitem[\protect\citeauthoryear{{Perna}, {Lanzuisi}, {Brusa}, {Cresci}  \&
  {Mignoli}}{{Perna} et~al.}{2017}]{Perna+17}
{Perna} M.,  {Lanzuisi} G.,  {Brusa} M.,  {Cresci} G.,   {Mignoli} M.,  2017,
  preprint, \href {http://adsabs.harvard.edu/abs/2017arXiv170508388P} {}
  (\mn@eprint {arXiv} {1705.08388})

\bibitem[\protect\citeauthoryear{{Prochaska}, {Kasen}  \& {Rubin}}{{Prochaska}
  et~al.}{2011}]{Prochaska+11}
{Prochaska} J.~X.,  {Kasen} D.,   {Rubin} K.,  2011, \mn@doi [\apj]
  {10.1088/0004-637X/734/1/24}, \href
  {http://adsabs.harvard.edu/abs/2011ApJ...734...24P} {734, 24}

\bibitem[\protect\citeauthoryear{{Rowlands}, {Wild}, {Nesvadba}, {Sibthorpe},
  {Mortier}, {Lehnert}  \& {da Cunha}}{{Rowlands} et~al.}{2015}]{Rowlands+15}
{Rowlands} K.,  {Wild} V.,  {Nesvadba} N.,  {Sibthorpe} B.,  {Mortier} A.,
  {Lehnert} M.,   {da Cunha} E.,  2015, \mn@doi [\mnras]
  {10.1093/mnras/stu2714}, \href
  {http://adsabs.harvard.edu/abs/2015MNRAS.448..258R} {448, 258}

\bibitem[\protect\citeauthoryear{{Rubin}, {Prochaska}, {Koo}, {Phillips},
  {Martin}  \& {Winstrom}}{{Rubin} et~al.}{2014}]{Rubin+14}
{Rubin} K.~H.~R.,  {Prochaska} J.~X.,  {Koo} D.~C.,  {Phillips} A.~C.,
  {Martin} C.~L.,   {Winstrom} L.~O.,  2014, \mn@doi [\apj]
  {10.1088/0004-637X/794/2/156}, \href
  {http://adsabs.harvard.edu/abs/2014ApJ...794..156R} {794, 156}

\bibitem[\protect\citeauthoryear{{Rupke} \& {Veilleux}}{{Rupke} \&
  {Veilleux}}{2013}]{Rupke+13}
{Rupke} D.~S.~N.,  {Veilleux} S.,  2013, \mn@doi [\apj]
  {10.1088/0004-637X/768/1/75}, \href
  {http://adsabs.harvard.edu/abs/2013ApJ...768...75R} {768, 75}

\bibitem[\protect\citeauthoryear{{Rupke} \& {Veilleux}}{{Rupke} \&
  {Veilleux}}{2015}]{Rupke+15}
{Rupke} D.~S.~N.,  {Veilleux} S.,  2015, \mn@doi [\apj]
  {10.1088/0004-637X/801/2/126}, \href
  {http://adsabs.harvard.edu/abs/2015ApJ...801..126R} {801, 126}

\bibitem[\protect\citeauthoryear{{Rupke}, {Veilleux}  \& {Sanders}}{{Rupke}
  et~al.}{2005a}]{Rupke+05a}
{Rupke} D.~S.,  {Veilleux} S.,   {Sanders} D.~B.,  2005a, \mn@doi [\apjs]
  {10.1086/432886}, \href {http://adsabs.harvard.edu/abs/2005ApJS..160...87R}
  {160, 87}

\bibitem[\protect\citeauthoryear{{Rupke}, {Veilleux}  \& {Sanders}}{{Rupke}
  et~al.}{2005b}]{Rupke+05b}
{Rupke} D.~S.,  {Veilleux} S.,   {Sanders} D.~B.,  2005b, \mn@doi [\apjs]
  {10.1086/432889}, \href {http://adsabs.harvard.edu/abs/2005ApJS..160..115R}
  {160, 115}

\bibitem[\protect\citeauthoryear{{Rupke}, {Veilleux}  \& {Sanders}}{{Rupke}
  et~al.}{2005c}]{Rupke+05c}
{Rupke} D.~S.,  {Veilleux} S.,   {Sanders} D.~B.,  2005c, \mn@doi [\apj]
  {10.1086/444451}, \href {http://adsabs.harvard.edu/abs/2005ApJ...632..751R}
  {632, 751}

\bibitem[\protect\citeauthoryear{{Sarzi}, {Kaviraj}, {Nedelchev}, {Tiffany},
  {Shabala}, {Deller}  \& {Middelberg}}{{Sarzi} et~al.}{2016}]{Sarzi+16}
{Sarzi} M.,  {Kaviraj} S.,  {Nedelchev} B.,  {Tiffany} J.,  {Shabala} S.~S.,
  {Deller} A.~T.,   {Middelberg} E.,  2016, \mn@doi [\mnras]
  {10.1093/mnrasl/slv165}, \href
  {http://adsabs.harvard.edu/abs/2016MNRAS.456L..25S} {456, L25}

\bibitem[\protect\citeauthoryear{{Sato}, {Martin}, {Noeske}, {Koo}  \&
  {Lotz}}{{Sato} et~al.}{2009}]{Sato+09}
{Sato} T.,  {Martin} C.~L.,  {Noeske} K.~G.,  {Koo} D.~C.,   {Lotz} J.~M.,
  2009, \mn@doi [\apj] {10.1088/0004-637X/696/1/214}, \href
  {http://adsabs.harvard.edu/abs/2009ApJ...696..214S} {696, 214}

\bibitem[\protect\citeauthoryear{{Savage} \& {Sembach}}{{Savage} \&
  {Sembach}}{1996}]{Savage+96}
{Savage} B.~D.,  {Sembach} K.~R.,  1996, \mn@doi [\araa]
  {10.1146/annurev.astro.34.1.279}, \href
  {http://adsabs.harvard.edu/abs/1996ARA%26A..34..279S} {34, 279}

\bibitem[\protect\citeauthoryear{{Sell} et~al.,}{{Sell} et~al.}{2014}]{Sell+14}
{Sell} P.~H.,  et~al., 2014, \mn@doi [\mnras] {10.1093/mnras/stu636}, \href
  {http://adsabs.harvard.edu/abs/2014MNRAS.441.3417S} {441, 3417}

\bibitem[\protect\citeauthoryear{{Snyder}, {Cox}, {Hayward}, {Hernquist}  \&
  {Jonsson}}{{Snyder} et~al.}{2011}]{Snyder+11}
{Snyder} G.~F.,  {Cox} T.~J.,  {Hayward} C.~C.,  {Hernquist} L.,   {Jonsson}
  P.,  2011, \mn@doi [\apj] {10.1088/0004-637X/741/2/77}, \href
  {http://adsabs.harvard.edu/abs/2011ApJ...741...77S} {741, 77}

\bibitem[\protect\citeauthoryear{{Tremonti} et~al.,}{{Tremonti}
  et~al.}{2004}]{Tremonti+04}
{Tremonti} C.~A.,  et~al., 2004, \mn@doi [\apj] {10.1086/423264}, \href
  {http://adsabs.harvard.edu/abs/2004ApJ...613..898T} {613, 898}

\bibitem[\protect\citeauthoryear{{Tremonti}, {Moustakas}  \&
  {Diamond-Stanic}}{{Tremonti} et~al.}{2007}]{Tremonti+07}
{Tremonti} C.~A.,  {Moustakas} J.,   {Diamond-Stanic} A.~M.,  2007, \mn@doi
  [\apjl] {10.1086/520083}, \href
  {http://adsabs.harvard.edu/abs/2007ApJ...663L..77T} {663, L77}

\bibitem[\protect\citeauthoryear{{Tripp} et~al.,}{{Tripp}
  et~al.}{2011}]{Tripp+11}
{Tripp} T.~M.,  et~al., 2011, \mn@doi [Science] {10.1126/science.1209850},
  \href {http://adsabs.harvard.edu/abs/2011Sci...334..952T} {334, 952}

\bibitem[\protect\citeauthoryear{{Vazdekis}, {S{\'a}nchez-Bl{\'a}zquez},
  {Falc{\'o}n-Barroso}, {Cenarro}, {Beasley}, {Cardiel}, {Gorgas}  \&
  {Peletier}}{{Vazdekis} et~al.}{2010}]{Vazdekis+10}
{Vazdekis} A.,  {S{\'a}nchez-Bl{\'a}zquez} P.,  {Falc{\'o}n-Barroso} J.,
  {Cenarro} A.~J.,  {Beasley} M.~A.,  {Cardiel} N.,  {Gorgas} J.,   {Peletier}
  R.~F.,  2010, \mn@doi [\mnras] {10.1111/j.1365-2966.2010.16407.x}, \href
  {http://adsabs.harvard.edu/abs/2010MNRAS.404.1639V} {404, 1639}

\bibitem[\protect\citeauthoryear{{Weiner} et~al.,}{{Weiner}
  et~al.}{2009}]{Weiner+09}
{Weiner} B.~J.,  et~al., 2009, \mn@doi [\apj] {10.1088/0004-637X/692/1/187},
  \href {http://adsabs.harvard.edu/abs/2009ApJ...692..187W} {692, 187}

\bibitem[\protect\citeauthoryear{{Wild}, {Walcher}, {Johansson}, {Tresse},
  {Charlot}, {Pollo}, {Le F{\`e}vre}  \& {de Ravel}}{{Wild}
  et~al.}{2009}]{Wild+09}
{Wild} V.,  {Walcher} C.~J.,  {Johansson} P.~H.,  {Tresse} L.,  {Charlot} S.,
  {Pollo} A.,  {Le F{\`e}vre} O.,   {de Ravel} L.,  2009, \mn@doi [\mnras]
  {10.1111/j.1365-2966.2009.14537.x}, \href
  {http://adsabs.harvard.edu/abs/2009MNRAS.395..144W} {395, 144}

\bibitem[\protect\citeauthoryear{{Wild}, {Almaini}, {Dunlop}, {Simpson},
  {Rowlands}, {Bowler}, {Maltby}  \& {McLure}}{{Wild} et~al.}{2016}]{Wild+16}
{Wild} V.,  {Almaini} O.,  {Dunlop} J.,  {Simpson} C.,  {Rowlands} K.,
  {Bowler} R.,  {Maltby} D.,   {McLure} R.,  2016, \mn@doi [\mnras]
  {10.1093/mnras/stw1996}, \href
  {http://adsabs.harvard.edu/abs/2016MNRAS.463..832W} {463, 832}

\bibitem[\protect\citeauthoryear{{Wong} et~al.,}{{Wong} et~al.}{2012}]{Wong+12}
{Wong} O.~I.,  et~al., 2012, \mn@doi [\mnras]
  {10.1111/j.1365-2966.2011.20159.x}, \href
  {http://adsabs.harvard.edu/abs/2012MNRAS.420.1684W} {420, 1684}

\bibitem[\protect\citeauthoryear{{Yesuf}, {Faber}, {Trump}, {Koo}, {Fang},
  {Liu}, {Wild}  \& {Hayward}}{{Yesuf} et~al.}{2014}]{Yesuf+14}
{Yesuf} H.~M.,  {Faber} S.~M.,  {Trump} J.~R.,  {Koo} D.~C.,  {Fang} J.~J.,
  {Liu} F.~S.,  {Wild} V.,   {Hayward} C.~C.,  2014, \mn@doi [\apj]
  {10.1088/0004-637X/792/2/84}, \href
  {http://adsabs.harvard.edu/abs/2014ApJ...792...84Y} {792, 84}

\bibitem[\protect\citeauthoryear{{Yesuf}, {French}, {Faber}  \& {Koo}}{{Yesuf}
  et~al.}{2017b}]{Yesuf+17b}
{Yesuf} H.~M.,  {French} K.~D.,  {Faber} S.~M.,   {Koo} D.~C.,  2017b,
  preprint, \href {http://adsabs.harvard.edu/abs/2017arXiv170500668Y} {}
  (\mn@eprint {arXiv} {1705.00668})

\bibitem[\protect\citeauthoryear{{Yesuf} et~al.,}{{Yesuf}
  et~al.}{2017a}]{Yesuf+17a}
{Yesuf} H.~M.,  et~al., 2017a, preprint, \href
  {http://adsabs.harvard.edu/abs/2017arXiv170408348Y} {} (\mn@eprint {arXiv}
  {1704.08348})

\bibitem[\protect\citeauthoryear{{Zakamska} \& {Greene}}{{Zakamska} \&
  {Greene}}{2014}]{Zakamska+14}
{Zakamska} N.~L.,  {Greene} J.~E.,  2014, \mn@doi [\mnras]
  {10.1093/mnras/stu842}, \href
  {http://adsabs.harvard.edu/abs/2014MNRAS.442..784Z} {442, 784}

\bibitem[\protect\citeauthoryear{{Zwaan}, {Kuntschner}, {Pracy}  \&
  {Couch}}{{Zwaan} et~al.}{2013}]{Zwaan+13}
{Zwaan} M.~A.,  {Kuntschner} H.,  {Pracy} M.~B.,   {Couch} W.~J.,  2013,
  \mn@doi [\mnras] {10.1093/mnras/stt496}, \href
  {http://adsabs.harvard.edu/abs/2013MNRAS.432..492Z} {432, 492}

\makeatother
\end{thebibliography}



\section{Appendix}

\subsection{Wind in H$\delta > 6$\,{\AA} AGN PSBs}

\begin{figure}
\centering
\includegraphics[width=0.9\linewidth]{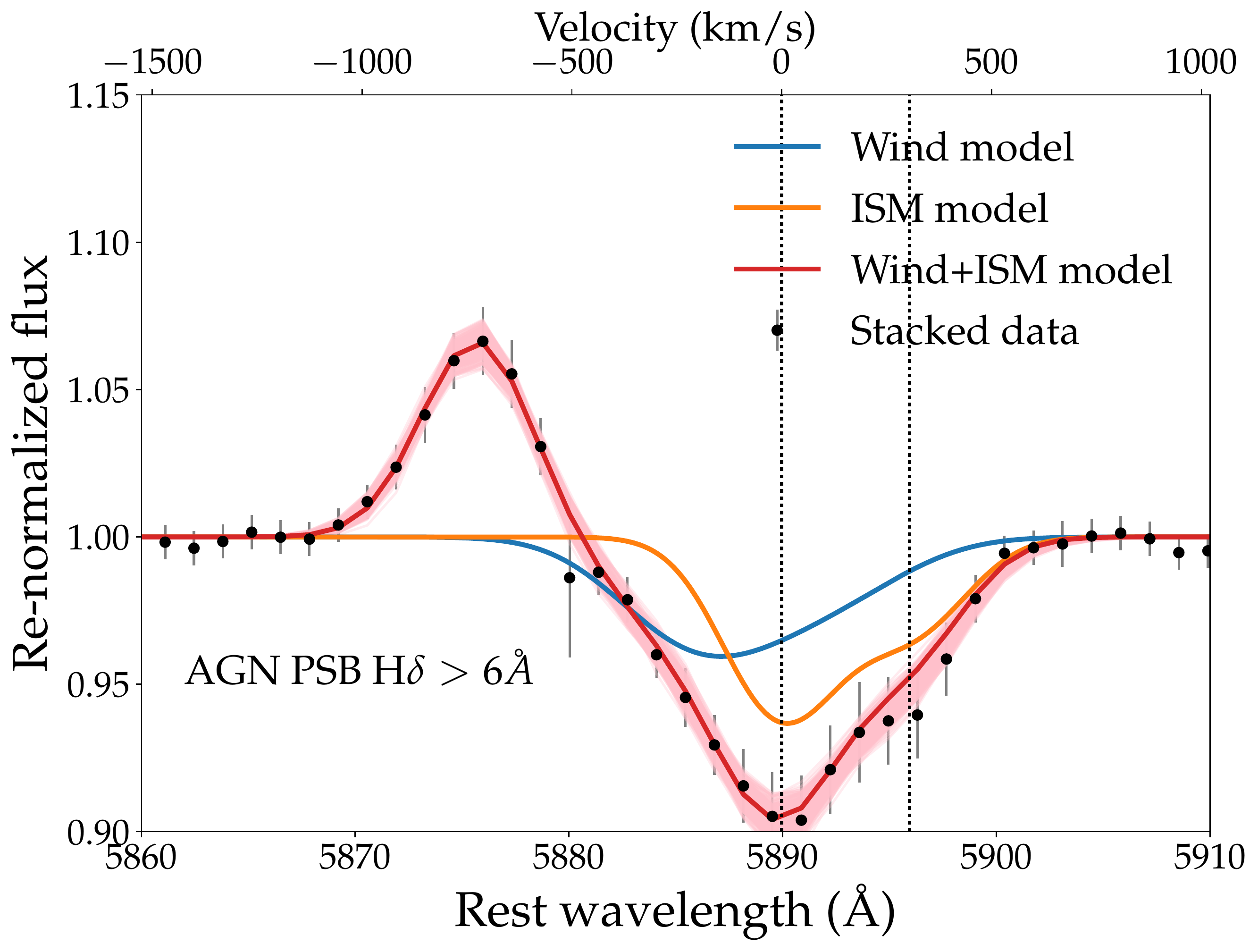}
\caption{The \ion{Na}{i} profile fit for the AGN PSBs with H$\delta > 6$\,{\AA}. The estimated centroid and maximum velocities are $-197^{+64}_{-65} $and $-556^{+67}_{-76}$\,\kms respectively. The equivalent width of the ISM component is $0.6^{+0.1}_{-0.2}${\AA} while that of the wind component is $0.5^{+0.2}_{-0.1}${\AA}. Regardless of how the AGN PSBs are defined the winds in these galaxies are weak. \label{fig:psb_hd6_fit}}
\end{figure}

The clustering of points around H$\delta > 5$\,{\AA}, in Figure~\ref{fig:hd_ha}, suggests some of the AGN PSB candidates that overlap with normal star-forming galaxies, may not be (strong) PSBs. In Figure~\ref{fig:psb_hd6_fit}, we show that if we restrict our AGN PSB sample to those with H$\delta > 6$\,{\AA}, we get similar wind velocities and strength presented in section~\ref{sec:analy}.

\subsection{Winds in H$\delta > 5$\,{\AA} Control Sample}

\begin{figure}
\centering
\includegraphics[width=0.9\linewidth]{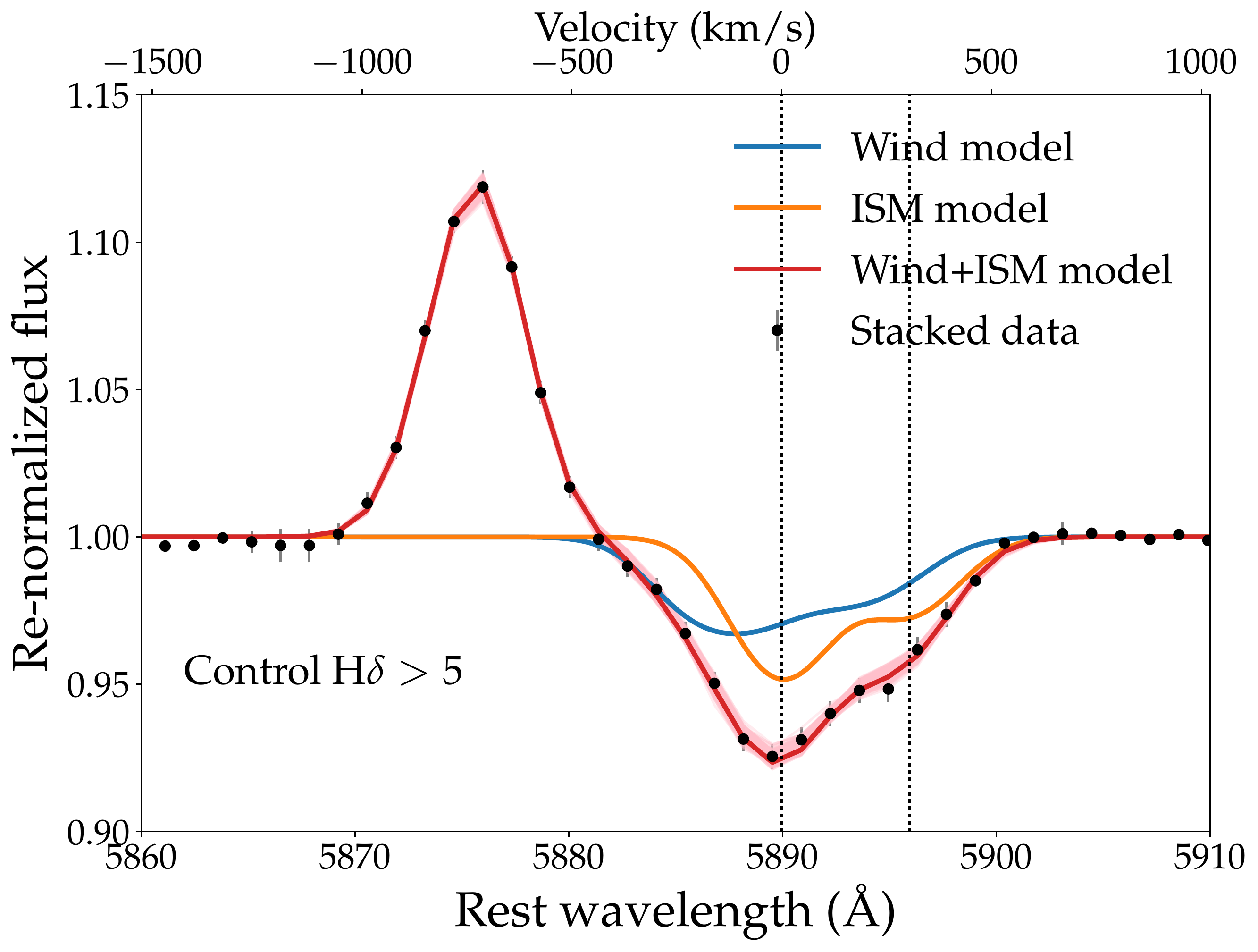}
\caption{The \ion{Na}{i} profile fit for the control sample of AGN PSBs. Unlike the corresponding control sample in the main text, the plotted control sample has H$\delta > 5$\,{\AA}. The estimated centroid and maximum velocities are $-121^{+22}_{-22} $and $-377^{+21}_{-24}$\,\kms respectively. The equivalent width of the ISM component is $0.42^{+0.05}_{-0.06}$\,{\AA} while that of the wind component is $0.37^{+0.06}_{-0.05}$\,{\AA}. The wind velocity, the wind and ISM equivalent width (EW) are similar to the control sample in the main text. This result indicates that the difference in wind velocities between the AGN PSBs and the control sample is due to AGN and is not likely due to a starburst related effects. \label{fig:hd5_fit}}
\end{figure}

AGN PSBs are defined to have H$\delta$ > 5\,{\AA}. But in our control sample selection criteria, we have also included control galaxies with H$\delta$ > 5\,{\AA}. Perhaps one may suspect that galaxies with H$\delta$ > 5\,{\AA} may have unusual properties unrelated to AGN that affect their \ion{Na}{i} profiles. In this section, we confirm that the wind properties of the control sample with H$\delta$ > 5\,{\AA} are similar to that of the control sample presented in the main text. Therefore, for galaxies that have H$\delta$ > 5\,{\AA} (i.e., PSBs) having AGN in the galaxies boosts the mean and the maximum velocities by a factor of 2. But the AGN winds are still weak. Figure~\ref{fig:hd5_fit} shows the \ion{Na}{i} profile fitting results of the control sample with H$\delta$ > 5\,{\AA}.

\bsp	
\label{lastpage}
\end{document}